\documentclass[fleqn,usenatbib]{mnras}

\usepackage{amsmath} 

\usepackage[oT2,T1]{fontenc}
\usepackage{xspace}
\usepackage{ae,aecompl}
\usepackage{hyperref}

\usepackage{enumitem} 

\usepackage{txfonts}

\usepackage{graphicx}	
\usepackage{multicol}   
\usepackage{multirow}   
\usepackage{pgfplots}
\usepackage{float} 
\usepackage[footnotesize]{caption}
\usepackage{colortbl}   
\usepackage{lscape}    
\usepackage{pdflscape} 
\usepackage{afterpage} 
\usepackage{diagbox} 
\definecolor{Mygrey}{rgb}{0.7,0.7,0.7}
\definecolor{Mylightgrey}{rgb}{0.9,0.9,0.9}
\definecolor{Myverylightgrey}{rgb}{0.97,0.97,0.97}
\definecolor{Myyellowgreen}{rgb}{0.5,0.5,0.2}
\definecolor{Myverydarkgrey}{rgb}{0.4,0.4,0.4}
\definecolor{Myred}{rgb}{0.7,0,0}
\definecolor{Mygreen}{rgb}{0,0.5,0.5}
\definecolor{Mydarkblue}{rgb}{0,0,0.5}
\definecolor{Mypurple}{rgb}{0.5,0,0.5}
\definecolor{Myblue}{rgb}{0,0,0.7}

\usepackage{amsbsy, amssymb, amsfonts}
\usepackage{bm}		

\usepackage{mathtools} 

\usepackage{algorithm}
\usepackage[noend]{algpseudocode}

\usepackage{xcolor} 

\usepackage{url}

\definecolor{gray65}{gray}{0.65} 
\definecolor{uclcolorlight}{RGB}{102,102,102}

\definecolor{NavyBlue}{RGB}{47,113,194}
\definecolor{greenmat}{RGB}{0,136,43}
\definecolor{org}{RGB}{222,106,16}
\definecolor{greenmat2}{RGB}{0,166,43}
\definecolor{org2}{RGB}{240,100,16}

\definecolor{lightgrey}{RGB}{230,230,230}
\definecolor{lightorg}{RGB}{255,220,180}
\definecolor{lighterorg}{RGB}{255,235,200}
\definecolor{lightblue}{RGB}{232,232,255}
\definecolor{lightgreen}{RGB}{230,255,230}
\definecolor{lightred}{RGB}{255,220,220}

\definecolor{darkblue}{RGB}{0,0,139}
\definecolor{darkblue2}{RGB}{10,40,140}

\definecolor{YW}{RGB}{160,160,0}
\definecolor{TC}{RGB}{200,60,120}

\newcommand{\bb}{\mathbb}
\newcommand{\bs}[1]{\boldsymbol{#1}}
\newcommand{\bsmat}[1]{\mathbf{#1}}
\newcommand{\bc}[1]{\boldsymbol{\mathcal #1}}
\newcommand{\cl}{\mathcal}

\newcommand{\wt}{\widetilde}

\newcommand{\st}{%
  \ifmmode
  \ {\rm s.t.}\ %
  \else%
  \emph{s.t.}\@\xspace%
  \fi%
}

\newcommand{\ie}{\emph{i.e.}, }
\newcommand{\eg}{\emph{e.g.}, }

\newcommand{\rmh}{\mathrm{h}}
\newcommand{\rmd}{\mathrm{d}}

\newcommand{\rmr}{\mathrm{r}}

\newcommand{\Rbb}{\bb{R}} 
\newcommand{\Cbb}{\bb{C}} 

\DeclareMathOperator{\vect}{vec} 
\DeclareMathOperator{\Id}{{\rm \bsmat I}} 
\newcommand{\tinv}[1]{{\textstyle\frac{1}{#1}}}

\newcommand{\im}{\mathrm{i}\mkern1mu} 
\newcommand{\real}[1]{\rm{Re}{\left\{#1\right\}}} 
\newcommand{\conj}[1]{#1^{*}} 

\newcommand{\proba}[2][]{
  {
    \ifx&#1& \bb{P}\left[#2\right] 
    \else\bb{P} #1[#2 #1]\fi
  }
}
\newcommand{\E}[2][]{\ifthenelse{ \equal{#2}{} }{\bb{E}_{#1}}{\bb{E}_{#1}\left[#2\right]}} 
\newcommand{\iid}{%
  \ifmmode
  \mathrm{i.i.d.}%
  \else%
  i.i.d.\@\xspace%
  \fi%
}

\newcommand{\ud}{\mathrm{d}} 
\DeclareMathOperator{\diag}{diag}
\newcommand{\transpose}[1]{#1^\top} 
\newcommand{\norm}[3][]{#1\lVert#2#1\rVert_{#3}} 
\newcommand{\scp}[3][]{#1\langle #2, #3 #1\rangle} 

\DeclareMathOperator*{\argmin}{arg\,min}

\newcommand{\upto}[1]{\llbracket #1 \rrbracket} 

\newcommand{\RIPto}{\text{RIP}_{\ell_2/\ell_1}}

\newcommand{\intMa}[1][]{
   \ifthenelse{ \equal{#1}{} }{\bs{\cl I}}{\cl I_{#1}}
}
\newcommand{\intMOm}[2][]{
   \ifthenelse{ \equal{#2}{} }{\bs{\cl I}_\Omega}{\bs{\cl I}_\Omega #1[ #2 #1]} 
}
\newcommand{\intMOmT}[2][]{
   \ifthenelse{ \equal{#2}{} }{\bs{\cl I}_\Omega^*}{\bs{\cl I}_\Omega^* #1[ #2 #1]} 
}
\newcommand{\intMOmb}[2][]{
   \ifthenelse{ \equal{#2}{} }{\bc I_{\Omega_b}}{\bc I_{\Omega_b} #1[ #2 #1]} 
}
\newcommand{\intMOmt}[2][]{
   \ifthenelse{ \equal{#2}{} }{\bs{\cl I}_{\Omega(t)}}{\bs{\cl I}_{\Omega(t)} #1[ #2 #1]} 
}
\newcommand{\dintMOm}[2][]{
   \ifthenelse{ \equal{#2}{} }{\tilde{\bs{\cl I}}_\Omega}{\tilde{\bs{\cl I}}_\Omega #1[ #2 #1]} 
}

\newcommand{\ts}{\textstyle}

\newcommand{\sigvign}{\sigma^\circ}
\newcommand{\btoB}{_{b=1}^B}
\newcommand{\acqop}{\bsmat\Psi}
\newcommand{\acqopn}{\wt{\bsmat\Psi}}
\newcommand{\imop}{\bsmat\Phi}

\newcommand{\noisemat}{\bsmat\Sigma_{\rm n}}
\newcommand{\covmat}[1][]{
   \ifthenelse{ \equal{#1}{} }{\bsmat{C}}{C_{#1}} 
}
\newcommand{\sampcovmat}[1][]{
   \ifthenelse{ \equal{#1}{} }{\bsmat C}{C_{#1}} 
}

\newcommand{\ddirty}{\bs\sigma_\rmd}
\DeclareMathOperator{\rlog}{rlog}

\renewcommand{\leq}{\leqslant}
\renewcommand{\geq}{\geqslant}

\usepackage[normalem]{ulem} 

\newcommand{\replace}[2][\@empty]
{ \ifbool{replace}
  {
  #2
  }
  {
  \ifx\@empty#1\relax \textcolor{greenmat2}{[BY]: #2} 
  \else\textcolor{org2}{[REPLACE]: #1} \\ \textcolor{greenmat2}{[BY]: #2}\fi
  }
}
\newcommand{\br}[1]{\textcolor{red}{[\textbf{BR:#1}]}}
\renewcommand{\br}[1]{}

\title[MROP]{MROP: Modulated Rank-One Projections for compressive radio interferometric imaging}

\author[O. Leblanc, C. S. Chu, L. Jacques, and Y. Wiaux]{
Olivier Leblanc$^{1}$
\thanks{Equal contribution.}, 
Chung San Chu$^{2}$\textcolor{blue}{\footnotemark[1]},
Laurent Jacques$^{1}$,
and Yves Wiaux$^{2}$\thanks{E-mail: \href{mailto:y.wiaux@hw.ac.uk}{y.wiaux@hw.ac.uk}}
\\
$^{1}$ISPGroup, INMA/ICTEAM, UCLouvain, 1348 Louvain-la-Neuve, Belgium.\\
$^{2}$Institute of Sensors, Signals and Systems, Heriot-Watt University, Edinburgh EH14 4AS, UK.\\
}
\date{Last updated 2025 January 26; in original form 2025 January 26}

\pubyear{2025}

\begin{document}
\label{firstpage}
\pagerange{\pageref{firstpage}--\pageref{lastpage}}
\maketitle

\begin{abstract}
    The emerging generation of radio-interferometric (RI) arrays are set to form images of the sky with a new regime of sensitivity and resolution. This implies a significant increase in visibility data volumes, which for single-frequency observations will scale as $\cl O(Q^2B)$ for $Q$ antennas and $B$ short-time integration intervals (or batches), calling for efficient data dimensionality reduction techniques. This paper proposes a new approach to data compression during acquisition, coined \emph{modulated rank-one projection} (MROP). MROP compresses the $Q\times Q$ batchwise covariance matrix into a smaller number $P$ of random rank-one projections and compresses across time by trading $B$ for a smaller number $M$ of random modulations of the ROP measurement vectors.
    Firstly, we introduce a dual perspective on the MROP acquisition, which can either be understood as random beamforming, or as a post-correlation compression. 
    Secondly, we analyse the noise statistics of MROPs and demonstrate that the random projections induce a uniform noise level across measurements independently of the visibility-weighting scheme used. Thirdly, we propose a detailed analysis of the memory and computational cost requirements across the data acquisition and image reconstruction stages, with comparison to state-of-the-art dimensionality reduction approaches. 
    Finally, the MROP model is validated for monochromatic intensity imaging both in simulation and from real data, with comparison to the classical and \emph{baseline-dependent averaging} (BDA) models, and using the uSARA optimisation algorithm for image formation.  Our results suggest that the data size necessary to preserve imaging quality using MROPs is reduced to the order of image size, well below the original and BDA data sizes.
\end{abstract}

\begin{keywords}
techniques: image processing -- techniques: interferometric.
\end{keywords}


\section{Introduction}
\label{sec:intro}

Data acquisition by interferometry in radio astronomy relies on an array of strategically placed antennas measuring, through the signal covariance matrix, a noisy and nonuniform partial coverage of the spatial Fourier domain of the image of interest, \ie a small portion of the celestial sphere accessible to the instrument. 
Radio interferometry (RI) is fundamental in studying a wide range of astronomical phenomena, including star formation, galaxy evolution, and the cosmic microwave background.

The classical model stores in memory all the visibilities---the Fourier samples obtained from the interferometric acquisition.
While current arrays like \emph{Very Large Array} (VLA) \citep{thompson80vla} or MeerKAT \citep{asad21meerkat} respectively have 27 and 64 antennas, future arrays such as the \emph{Square Kilometer Array} (SKA) are coming up with an increasing number of antennas.
SKA itself is to form petabyte-scale ($10^{15}$ bytes) images from exabyte-scale ($10^{18}$ bytes) data volumes \citep{scaife20}. 
Compression is thus emerging as a critical need to reduce the data size and make RI imaging scalable. 

Recently, \citet{leblanc2024b} proposed a compressive sensing technique applied directly at the level of the antenna measurements. It showed that \emph{beamforming}---a common technique of dephasing antenna signals---usually used to focus some region of the sky, is equivalent to sensing a rank-one projection (ROP) of the signal covariance matrix. Random beamforming was combined with random modulations across time to build the modulated rank-one projection (MROP) model that replaces the enormous number of visibilities with a tunable number of MROPs.
In \citet{leblanc2024b}, the emphasis was made on the mathematical foundations of the random beamforming. Image recovery guarantees were provided for image reconstruction under a simplistic model assuming sparsity in the image domain. 
The associated reduced sample complexity, theoretically derived in a simpler case without modulations and numerically obtained in phase transition diagrams, was shown to scale as $\cl O(K)$ where $K$ is the image sparsity. 

This paper significantly extends the study of the MROP model while still focussing on monochromatic intensity imaging from single-frequency observations.
Firstly, we discuss two acquisition schemes. The antenna-based \emph{random beamforming} acquisition scales linearly with the number of antennas, while the visibility-based acquisition is closer to the classical correlation computation and facilitates visibility weighting.
Secondly, the noise statistics of MROPs are derived.
Thirdly, the memory requirements and the computational costs of the acquisition and imaging steps are derived and compared with other models in the literature.
Finally, the reconstruction of images with high dynamic range in a realistic numerical setting as well as from real data demonstrates the potential of the MROP model. 

\subsection{Related Work} \label{sec:related}

Several compression techniques for RI that are connected to our method exist in the literature.
Among them, compressing the observation vector that collects all the visibilities by multiplication with a Gaussian random matrix (with much fewer rows than columns) was discussed in \citet{kartik17} and shown to increase the computational cost of the forward model significantly. \citet{kartik17} proposed an efficient Fourier reduction model approximating the optimal (in the least-squares sense) dimensionality reduction that projects the data with the adjoint of $\bs U$ which contains the left singular vectors of the SVD of the classical imaging operator $\bs\Phi:=\bs U \bs\Sigma\bs V^*$. 
On the other hand, \emph{baseline-dependent averaging} (BDA) \citep{wijnholds18,atemkeng18} offers an effective reduction in data volume by averaging consecutive visibilities corresponding to short baselines, \ie those with nearly constant frequency locations.  
Finally, standard visibility gridding, which consists of storing the \emph{dirty image}, is an efficient data dimensionality reduction technique when the original visibility data size is larger than the image size, as discussed in \citet{kartik17}. 
The dirty map can even be computed in an antenna-based way which reduces its acquisition cost \citep{thyagarajan17,krishnan23}. 
\emph{Random beamforming} for RI has first been suggested in \citet{aecal15}. 
Then, the connection of random beamforming to ROPs of the signal covariance matrix has been highlighted in \citet{leblanc2024b}, leveraging the theory of recovery guarantees via ROP to derive the sample complexity in a simplified case. 
This paper studies image reconstruction using the uSARA algorithm \citep{repetti20, terris22}, which is a sophisticated version of the proximal gradient method \citep{Beck_book}.

\subsection{Contributions} \label{sec:contributions}

In this work, we significantly increase the practicality of the MROP approach proposed by \citet{leblanc2024b} to RI by providing five novel contributions, all studied in the context of single-frequency monochromatic intensity imaging.

\paragraph*{Acquisition of MROPs.}
Two perspectives on acquiring MROPs are discussed in Sections~\ref{sec:mrop1} and \ref{sec:mrop2} on the grounds of their acquisition costs, their compatibilities with the existing correlators, and their compatibilities with visibility weighting. 
The first perspective, involving \emph{random beamforming}, scales linearly with the number of antennas but is shown to raise challenges for visibility weighting because the visibilities are never accessed. 
A possible solution that decomposes the matrix of weights as a sum of rank-one matrices is proposed in Section~\ref{sec:mrop_weighting}. 
The second perspective, coined \emph{visibility-based} acquisition of MROP, applies post-correlation compression so accesses the visibilities, hence enabling their weighting, and is also compatible with the existing correlators. 

\paragraph*{Noise statistics on MROPs.}
In Section~\ref{sec:mrop_noise}, we conduct a statistical analysis showing that MROP computation preserves the additive ~\iid Gaussian nature of noise in the visibilities, merely rescaling it, irrespective of the visibility weighting scheme.

\paragraph*{Reduced memory requirements.} The MROP model is shown in Section~\ref{sec:complexity} to benefit from reduced intermediate and final memory storage for the acquisition and imaging processes compared to the classical model, and for a similar computational cost.
The computational complexities and memory requirements of multiple models are synthesised in Table~\ref{tab:acquisition} for the acquisition and in Table~\ref{tab:imaging} for the imaging. 

\paragraph*{Validation on simulated data.} The performance of the MROP imaging model is numerically studied in Section~\ref{sec:setting} in various regimes, including four ground-truth images set to different values of dynamic range, each coupled with two antenna arrays (namely VLA and MeerKAT) and twenty five different $uv$-coverages, and a range of ratios between the image resolution, the number of visibilities, and the number of MROPs. 
The MROP model is compared with both the classical model and the computationally effective BDA model \citep{wijnholds18} using uniform visibility weighting and the uSARA reconstruction algorithm. It is shown that the MROP model features a high compression factor and good reconstruction performances at a computational cost similar to the one of the classical model.

\paragraph*{alidation on real data.}
Section~\ref{sec:real_data} explores the image reconstruction performance of the MROP model on real data.
Using uniform visibility weighting and the uSARA reconstruction algorithm on VLA observations, the MROP model demonstrates a high compression factor while forming an reconstructed image whose structures, dynamic range and data fidelity are consistent with those of the classical model.

\subsection{Notations and Conventions} \label{sec:notation}
 
The acquired data will be noted in vector form (\eg $\bs v$ and the same data computed from the imaging model will be overlined (\eg $\bar{\bs v}$). 
When elevating a vector $\bs v$ to some power $p$ as $\bs v^p$, this is to be understood elementwise.
The identity matrix of size $n \times n$ is denoted by $\Id_n$, and the vector of length $n$ with all values of 0 (resp. 1) is denoted by $\bs 0_n$ (resp. $\bs 1_n$). When it is clear from context, the subscript $n$ will be omitted.
The set of $Q \times Q$ Hermitian matrices in $\bb C^{Q \times Q}$ is denoted by $\cl H^Q$.
The set of index components is $\upto M := \{1,\ldots,M\}$; $\{s_q\}_{q=1}^Q$ is the set $\{s_1, \ldots, s_Q\}$, and $(a_q)_{q=1}^Q$ is the vector $(a_1,\ldots,a_Q)^\top$. 
The notations 
$\simeq$,
$\transpose{\cdot}$, 
$\conj{\cdot}$,
$\odot$,
$\times$,
$\scp{\cdot}{\cdot}$, 
$\otimes$,
$\bc T$,
correspond to 
approximately equal,
transpose, 
conjugate transpose, 
elementwise product, 
cross product, 
inner product, 
Kronecker product, 
and upper triangular matrix selection.
The $p$-norm (or $\ell_p$-norm) is $\norm{\bs x}{p}:=  (\sum_{i=1}^n |x_i|^p)^{1/p}$ for $\bs x \in \bb C^n$ and $p\geq 1$. 
Given $\bsmat A \in \bb C^{n\times n}$, $\bs a \in \bb C^n$, the operator $\diag(\bsmat A) \in \bb C^n$ extracts the diagonal of $\bsmat A$, $\diag(\bs a) \in \bb C^{n\times n}$ is the diagonal matrix such that $\diag(\bs a)_{ii} = a_i$, 
$\norm{\bsmat A}{\rm F}:=\sqrt{\sum_{i,j=1}^n |A_{ij}|^2}$ is the Frobenius norm of $\bsmat A$,
and $\|\bsmat A\|_{\rm S}:=\underset{\norm{\bs x}{2}= 1}{\max}~ \norm{\bsmat A \bs x}{2}$ is its operator norm.  
The indicator function for a set $\cl S$ is defined as $\iota_{\cl S}(x) :=  0$, if $x \in \cl S$ and $\infty$, otherwise.
The direct and inverse continuous Fourier transforms in $d$ dimensions (with $d \in \{1,2\}$) are defined by $\cl F[g](\bs k) := \int_{\bb R^d} g(\bs s) e^{-\im 2\pi \bs k^\top \bs s} \ud \bs s$, with $g: \bb R^d \to \bb C^d$, $\bs k \in \bb R^d$, and $\cl F^{-1}[\hat g](\bs s) = 
\int_{\bb R^d} \hat{g}(\bs k) e^{ \im 2\pi \bs k^\top \bs s} \ud \bs k$. 
\section{Classical Model} \label{sec:classic}

This section revisits the classical approach for acquiring visibilities from the covariance matrix of the antenna measurement vector, where each visibility is assigned a specific weight.

\subsection{From the Antenna Signals to the Visibilities}

\begin{figure}
\centering 
\includegraphics[width=\linewidth]{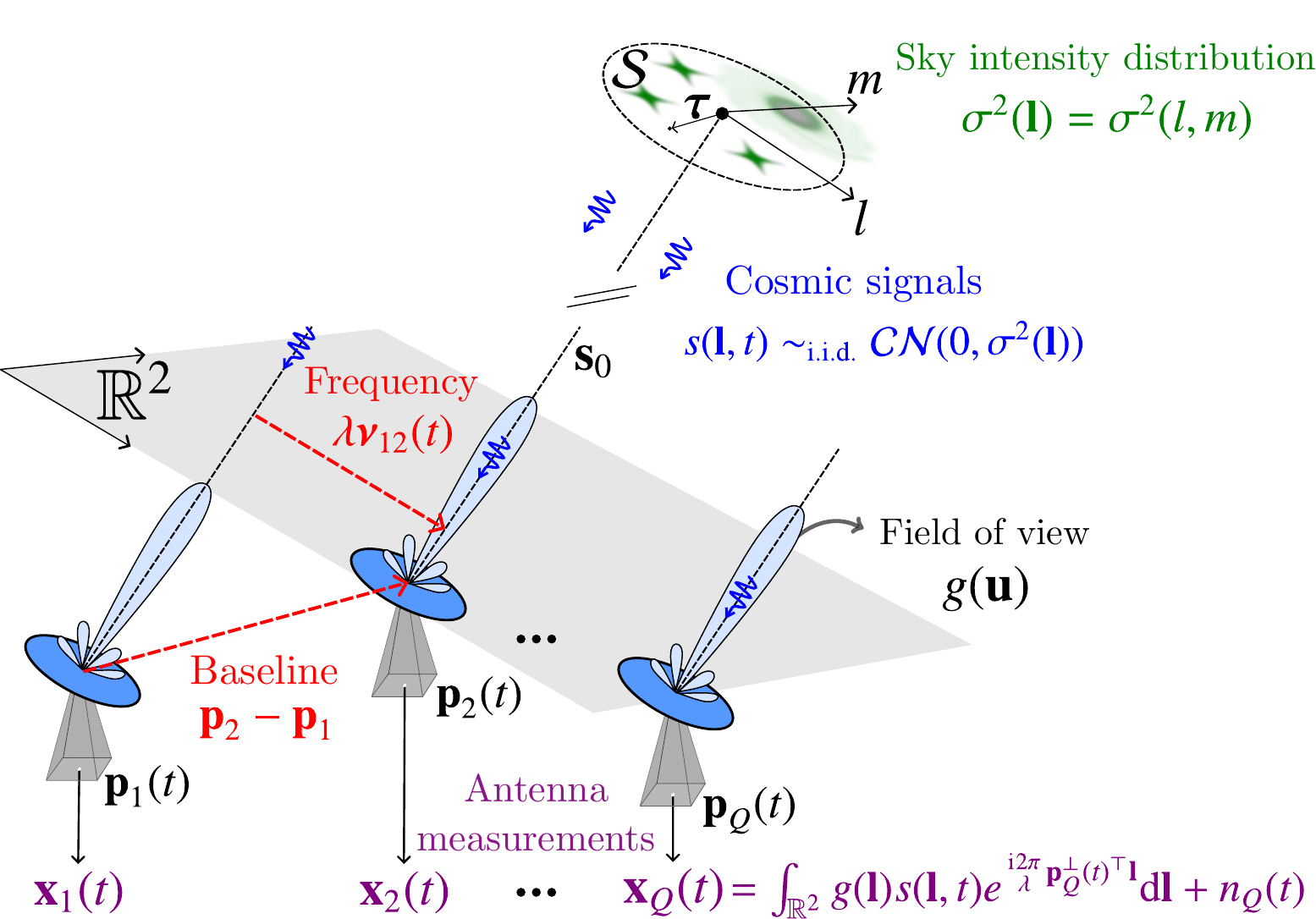}
\caption{Schematic of the radio interferometric sensing context. Far-away cosmic electric fields following a Gaussian random process, \ie $s(\bs l,t)\sim_{\iid} \cl C\cl N(0,\sigma^2(\bs l))$ with an intensity distribution $\sigma^2(\bs l)$, are received by $Q$ antennas. The antennas, with projected positions in $\Omega(t):=\{ \bs p_q^\perp(t) \}_{q=1}^Q$, have the same direction-dependent gain $g(\bs l)$ focusing on a specific region $\cl S$ of the sky. Each antenna $q\in \upto{Q}$ integrates---with its own geometric delay $\bs p_q^\perp(t)^\top\bs l$ the electric field contributions from all direction into a noisy measurement $x_q(t)=\int_{\Rbb^2} g(\bs l)s(\bs l,t)e^{\frac{\im2\pi}{\lambda}\bs p_q^\perp(t)^\top\bs l} \ud\bs l + n_q(t)$.} 
\label{fig:schematic}
\end{figure}

Let us consider the context of radio astronomical single-frequency monochromatic intensity imaging depicted in Fig.~\ref{fig:schematic} with $Q$ antennas receiving a linearly-polarised complex Gaussian electric field of amplitude $s(\bs l,t) \sim_\iid \cl C\cl N(0,\sigma^2(\bs l))$ (with the assumption $\bb E [s(\bs l,t) s^*(\bs l',t)] = \sigma^2(\bs l) \delta(\bs l-\bs l')$ \citep{vancittert34}) from the sky with power flux density distribution $\sigma^2(\bs l)$ [W/m$^2$]---the monochromatic intensity image of interest. We write $\bs l=(l,m)$ the direction cosines of the portion of the sky looked from the array formed by the antennas. 
More precisely, a \emph{direction cosine} coordinate system fixed with its origin at the center of the Earth is chosen such that the center of the field to be imaged is denoted by the unit vector $\bs s_0 = (l,m,n) = (0,0,1)$. The other directions in the region of interest are denoted by $\bs s=\bs s_0+\bs\tau$ with $\bs\tau = (l,m,\sqrt{1-l^2-m^2})$. 
The region of interest will be considered sufficiently small to be approximated as a plane, or equivalently $\sqrt{1-l^2-m^2}\simeq 1$. The set $\Omega(t):=\{\bs p_q^\perp(t)\}_{q=1}^Q \subset \Rbb^2$ denotes the projection onto the plane perpendicular to $\bs s_0$ of the instantaneous position of the $Q$ antennas---moving in time due to the rotation of the Earth.

In this context, the antenna measurement vector at time $t$ writes as $\bs x(t) = \bar{\bs x}(t) + \bs n(t)$.
The noiseless part of the $q$-th measurement writes
\begin{equation} \label{eq:31_meas}
    \bar x_q(t) = \int_{\Rbb^2} g(\bs l) s(\bs l,t) e^{\frac{\im 2\pi}{\lambda} \bs p_q^\perp(t)^\top \bs l} \ud\bs l,
\end{equation}
where $\lambda$ is the wavelength of the observed electric field and $g(\bs l)$ is the same direction-dependent gain for all antennas \citep{leblanc2024b}. 
The other term is a complex zero mean white Gaussian additive noise $\bs n(t) \sim_\iid \cl{CN}(\bs 0,\noisemat)$ where the noise covariance matrix $\noisemat$ is diagonal. 

Separating the total acquisition time interval into \emph{batches} indexed by $b$ and of duration $IT$ with $I$ samples per batch and a \emph{sampling period} $T$, we can define a \emph{signal set} $\cl X$ stacking all the antenna measurement vectors as 
\begin{equation} \label{eq:xb}
    \ts\cl X := \bigcup_{b \in \upto{B}} \cl X_b,\quad \cl X_b :=\{\bs x_b[i],~i\in\upto{I}\},  
\end{equation}
where $\bs x_b[i] = \bs x( (bI+i)T )$. 
Each batch $b$ has a sufficiently short duration to assume a stationary set of antenna projected positions $\Omega_b:=\{\bs p_{bq}^\perp\}_{q=1}^Q \subset \Rbb^2$ with the time averaging $\bs p_{bq} := \langle \bs p_q(t) \rangle_{t\in \cl T_b}$ where $\cl T_b := [bIT,(b+1)IT]$ is the time interval of batch $b$.

Leveraging the \emph{Van Cittert-Zernike} theorem \citep{vancittert34,zernike38} and assuming that the same realisation of the electric field $s(\bs l,t)$ is received simultaneously at time $t$ for all antennas, the classical acquisition scheme consists in computing, for each batch $b$, a \emph{sample covariance matrix} as
\begin{align} \label{eq:31_empi_covmat}
\begin{split}
    \sampcovmat_b(\cl X_b) &:= \frac{1}{I} \sum_{i=1}^I \bs x_b[i]\bs x_b^*[i] \\
    &\simeq \E[s,\bs n]{\bs x_b[i] \bs x_b[i]^*} = \intMOmb[]{\sigvign} + \noisemat,
\end{split}
\end{align}
where
\begin{equation} \label{eq:31_vign}
    (\intMOmb{\sigvign})_{jk} := \int_{\Rbb^2} \sigvign(\bs l) e^{\frac{-\im 2\pi}{\lambda} (\bs p_{bk}^\perp-\bs p_{bj}^\perp)^\top \bs l} \ud\bs l
\end{equation}
is the $jk$-th \emph{visibility}, \ie the $jk$-th entry of the $b$-th \emph{interferometric matrix} of the map $\sigvign$ where $\sigvign(\bs l):=g^2(\bs l)\sigma^2(\bs l)$ is the map \emph{vignetted} by (the square of) the antenna gain function (\ie the telescope's primary beam).

Overall, \eqref{eq:31_empi_covmat} and \eqref{eq:31_vign} show that RI corresponds to an interferometric system that is \emph{affine} in $\sigvign$. 
It is tantamount to sampling the 2-D Fourier transform of $\sigvign$ with a sampling pattern, or $uv$-coverage in each batch $b$ given by the difference multiset
\begin{equation} \label{eq:31_baselines}
\cl V_b := \tinv{\lambda}(\Omega_b - \Omega_b) = \{ \bs \nu_{bjk} := \tfrac{\bs p_{bj}^\perp - \bs p_{bk}^\perp}{\lambda} \}_{j,k=1}^Q,
\end{equation}
\ie $(\intMOmb{\sigvign})_{jk} = \cl F[\sigvign](\bs \nu_{bkj})$, then adding the diagonal covariance matrix of the noise $\noisemat$. The overall sampling pattern or $uv$-coverage results from the union of the coverages across batches: $\cl V := \bigcup\btoB \cl V_b$.

\subsection{Acquisition and Imaging Models}

In the classical model, two parts of each sample covariance matrix $\sampcovmat_b$ are abandoned: \emph{(i)} the diagonal containing $Q$ repetitions of the DC component $\cl F[\sigvign](\bs 0)$ plus the diagonal noise covariance matrix $\noisemat$, and \emph{(ii)} the lower triangular part of the sample covariance matrix $\sampcovmat_b$ which is identical to the upper part because it is Hermitian.
Consequently, within each batch $b \in \upto{B}$, the retained upper triangular part $\bc T\sampcovmat_b$, where the operator $\bc T$ zeroes all but the upper triangular of a matrix, contains $V:=Q(Q-1)/2$ visibilities, where $V$ is the number of antenna pairs, \ie \emph{baselines}, considered for correlation.
The identity $\bc T \sampcovmat_b \simeq \bc T \intMOmb{}[\sigma^\circ]$ shows that the acquisition extracts all the visibilities except the DC component.

The concatenation of the visibilities acquired over the $B$ batches and extracted from (the upper triangular part of) the interferometric matrices defined in \eqref{eq:31_vign} leads to the classical acquisition scheme 
\begin{equation} \label{eq:classic_acq}
    \bs v = \acqop(\cl X) + \bs\xi. 
\end{equation}
In \eqref{eq:classic_acq}, $\bs v \in \Cbb^{VB}$ contains the acquired visibilities, $\bs\Psi$ is the \emph{acquisition operator} defined as 
\begin{align}
\begin{split}
\label{eq:acqop}
    \cl X \mapsto\ & \acqop(\cl X):= (\acqop_1(\cl X_1)^\top,\ldots,\acqop_B(\cl X_B)^\top)^\top \in \Cbb^{VB},\\
    &\text{with}\ \acqop_b(\cl X_b) := \vect(\bc T\sampcovmat_b(\cl X_b))\in \Cbb^V,
\end{split}
\end{align} 
where $\bs\xi \sim \cl{CN}(\bs 0, \tau^2 \Id) \in \Cbb^{VB}$ is additive noise that accounts for statistical noise and other noise sources, such as quantum and quantisation noise \citep{thompson80vla}.
For simplicity and without loss of generality, we assume \iid Gaussian noise with constant noise level $\tau$ in our analysis.

Denoting the discretised vignetted map by $\bs\sigma \in \Rbb^{N}$, the classical RI forward imaging model \cite[Eq.~(2)]{onose16} associated with the acquisition model \eqref{eq:classic_acq} writes 
\begin{equation} \label{eq:classic}
\bar{\bs v} = \imop(\bs\sigma) + \bar{\bs\xi},
\end{equation}
where $\imop:=\bsmat{GFZ}$ 
is the \emph{imaging operator}, $\bar{\bs\xi} \in \Cbb^{VB}$ remains \iid additive noise sharing the same statistics as $\bs\xi$, \ie $\bar{\bs\xi} \sim \cl{CN}(\bs 0, \tau^2 \Id)$, and $\bsmat G \in \Cbb^{VB\times 4N}$ is a convolutional interpolation operator that interpolates the \emph{on-grid} frequencies obtained from the FFT $\bsmat{FZ}\bs\sigma$ in the 
continuous Fourier plane at frequencies corresponding to the difference sets $\{\cl V_b\}\btoB$ defined in \eqref{eq:31_baselines}, oversampled by the \emph{zero-padding} operator $\bsmat{Z} \in \{ 0,1\}^{4N\times N}$. 
This procedure is known as \emph{Non-Uniform Fast Fourier Transform} (NUFFT). 
The acquisition and imaging operators $\bs\Psi$ and $\bs\Phi$ are identical up to statistical noise, the origin of the approximation made in \eqref{eq:31_empi_covmat}, which means that 
\begin{equation} \label{eq:oplink}
    \bb E_{\cl X} \acqop(\cl X) = \imop(\bs\sigma).
\end{equation}

\subsection{Visibility Weighting} \label{sec:weighting}

\emph{Visibility weighting} is a common technique to improve the dynamic range and the ability to adjust the synthesised point spread function associated with the produced image \citep{briggs95}. 
Within each batch $b\in \upto{B}$, the visibilities can be weighted by an elementwise multiplication with the upper triangular part $\bc T \wt{\bsmat W}_b$ of a weighting matrix $\wt{\bsmat W}_b \in \cl H^{Q}$ as $\bc T \wt\sampcovmat_b := \bc T \wt{\bsmat W}_b \odot \bc T \sampcovmat_b$.
This modifies the classical acquisition and imaging models as
\begin{align}
    \bsmat W \bs v &= \bsmat W\acqop(\cl X) + \bsmat W \bs\xi \label{eq:classic_acq_w}, \\
    \bsmat W \bar{\bs v} &= \bsmat W\imop(\bs\sigma) + \bsmat W \bar{\bs\xi} \label{eq:classic_im_w},
\end{align}
where $\bsmat W \in \Rbb^{VB\times VB}$ is the diagonal visibility weighting operator defined as 
\begin{equation}
    \bs v = \begin{bmatrix} \bs v_1 \\ \vdots \\\bs v_B \end{bmatrix} \in \Cbb^{VB} \mapsto \bsmat W\bs v = \begin{bmatrix}
        \vect(\bc T \wt{\bsmat W}_1) \odot \bs v_1 \\ \vdots \\ \vect(\bc T \wt{\bsmat W}_b) \odot \bs v_B
    \end{bmatrix} \in \Cbb^{VB}.
\end{equation}
A schematic of the computations of this classical acquisition in \eqref{eq:classic_acq_w} is provided in Fig.~\ref{fig:acquisition}(a). 
Simple derivations show that visibility weighting modifies the noise statistics as $\bsmat{W}\bar{\bs\xi} \sim \cl{CN}(\bs 0, \tau^2 \diag(\bs w^2))$ with the vector of weights $\bs w := (\vect(\bc T \wt{\bsmat W}_1)^\top,\ldots,\vect(\bc T \wt{\bsmat W}_B)^\top)^\top$.

\emph{Natural} and \emph{uniform weighting} are two popular visibility weighting schemes \citep{briggs95}. 
For the classical scheme and for natural weighting, the weights are set as $\bs w_{\rm n}=\tau^{-1} \bs 1$ to ensure that the noise on the weighted visibilities is \iid with unit variance, \ie $\bsmat W_{\rm n}\bar{\bs\xi} \sim \cl{CN}(\bs 0, \Id)$. Uniform weighting additionally compensates for the density $\bs\rho \in \Rbb^{VB}$ of the visibilities in the $uv$ plane, \ie $\bs w_{\rm u}=\tau^{-1} \bs\rho^{-1/2}$, in order to precondition the sampling pattern and ease deconvolution. The resulting noise distribution reads as $\bsmat W_{\rm u}\bar{\bs\xi} \sim \cl{CN}(\bs 0, \diag(\bs\rho^{-1}))$, implying in particular that the noise on the weighted visibilities is not any more identically distributed. We explain in Section~\ref{sec:compression} how weighting is performed for each of the sensing models of interest, with random projection operators ensuring \iid noise on the final measurements, independent of the original choice of visibility weighting.
\section{Compression Models} \label{sec:compression}

This section introduces various compression models designed to reduce memory demands for both data storage and image reconstruction. After highlighting the limitations of existing post-correlation compression in Section~\ref{sec:post_corr}, the MROP model is presented as an advancement that overcomes these challenges. Two variants of the MROP acquisition model are detailed in Sections~\ref{sec:mrop1} and \ref{sec:mrop2}.

\subsection{Post-Correlation Compression} \label{sec:post_corr}

The issue with the classical scheme described in Section~\ref{sec:classic} is that the number of visibilities $VB=Q(Q-1)B/2$ scales as the square of the number of antennas $Q$, and it creates challenges in memory requirements and computational cost for image formation with just under a hundred antennas and thousands of batches. 
To reduce the memory requirements while still being able to estimate the image, several compression techniques have been considered in the past. 

\paragraph*{Dirty Image Model.}
For instance, applying the adjoint of the forward operator as $\ddirty:=\real{\acqop^* \bs v}$ and $\bar{\bs\sigma}_\rmd:=\real{\imop^* \bar{\bs v}}$ has the effect of \emph{gridding} the visibilities to form the \emph{dirty image} \citep{kartik17}. 
The associated acquisition and imaging models are
\begin{align} 
    \ddirty &= \real{\acqop^* \acqop(\cl X) + \acqop^* \bs\xi}, \\
    \bar{\bs\sigma}_\rmd &= \real{\imop^* \imop(\bs\sigma) + \imop^* \bar{\bs\xi}}, \label{eq:dirty_image_im}
\end{align}
with $\ddirty$, $ \bar{\bs\sigma}_\rmd \in \Rbb^N$, and $\acqop^*$ and $\imop^*$ are the adjoints of the acquisition and imaging operators, respectively.
This model reduces the memory requirements to $N$ (real) pixel values while only doubling the computational cost of one iteration of the classical forward model.
The visibility-weighted dirty image noise statistics can be shown to be $\imop^* \bsmat{W} \bar{\bs\xi} \sim \cl{CN}(\bs 0, \real{\imop^* \diag(\bs w^2) \imop}/2)$.
The noise in the dirty image pixels is not \iid but is correlated, which makes the image recovery more challenging.

\paragraph*{BDA.}
A second possibility named \emph{baseline-dependent averaging} (BDA) \citep{wijnholds18} consists in averaging the visibilities associated with low-frequency content over consecutive batches. 
This yields fewer measurements $VB_{\rm bda} \ll VB$ with an average number of batches per baseline $B_{\rm bda}:=\tinv{V}\sum_{i=1}^V B_i \ll B$ where $B_i$ is the reduced number of batches of the $i$-th pair of antennas. 
Applying the \emph{averaging operator} $\bs S \in \{1,1/2,1/3,1/4,\ldots\}^{VB_{\rm bda}\times VB}$, which contains a single entry per column but may have multiple values per row (summing to one) depending on the number of averaged visibilities, provides
\begin{align}
    \bs v_{\rm bda} &= \bsmat S \acqop(\cl X) + \bsmat S \bs\xi, \\
    \bar{\bs v}_{\rm bda} &= \imop_{\rm bda}(\bs\sigma) + \bsmat S \bar{\bs\xi},
    \label{eq:baseline_dep_av_im}
\end{align}
where $\imop_{\rm bda}(\bs\sigma) := \bsmat S \imop(\bs\sigma) =\bsmat G_{\rm bda}\bsmat{FZ}\bs\sigma$ and $\bsmat G_{\rm bda}:=\bsmat{SG} \in \Cbb^{VB_{\rm bda}\times 4N}$ is the averaged version of $\bsmat{G}$. 
An attractive feature of BDA is that the interpolation matrix $\bsmat{G}_{\rm bda}$ can be approximated directly from the $uv$-coverage chosen to represent the averaged visibilities, hence reducing the computational cost.

With BDA, the noise on the averaged visibilities is not \iid. Taking the multiplicity of each row of the averaging operator $\bsmat S$ yields the noise statistics $\bsmat S \bar{\bs\xi}~\sim~\cl{CN}(\bs 0, \tau^2 \diag(\bs s^{-1}))$, where the $i$-th component $s_i := \norm{\bsmat S_i}{0}$ of $\bs s$ counts the number of visibilities averaged in $\bar v_{{\rm bda}, i}$ for all $i \in \upto{VB_{\rm bda}}$.
The noise statistics of the averaged then weighted visibilities, with $\bsmat{W}_{\rm bda} \in \Rbb^{VB_{\rm bda}\times VB_{\rm bda}}$, is 
$\bsmat{W}_{\rm bda}\bsmat{S}\bar{\bs\xi} \sim \cl{CN}\big(\bs 0, \tau^2 \diag(\bs w_{\rm bda}^2 \odot \bs s^{-1})\big)$.
For BDA, natural weighting now applies different weights to the averaged visibilities depending on their noise level, \ie $\bs w_{\rm bda, \rm n} = \tau^{-1} \bs s^{1/2}$.
A uniform weighting of the averaged visibilities requires 
$\bs w_{\rm bda, \rm u} = \tau^{-1} \bs\rho_{\rm bda}^{-1/2} \odot \bs s^{1/2}$
where $\bs\rho_{\rm bda} \in \Rbb^{VB_{\rm bda}}$ is the density of the averaged visibilities in the $uv$ plane. 

BDA is cheap and can reduce the number of data points down to $VB_{\rm bda}/VB=0.1$ when also averaging over frequencies \citep{wijnholds18}, but the resulting compression does not leverage the small amount of information contained in low-complexity models of the image $\bs\sigma$. 

\paragraph*{Random Gaussian Compression.}
A third possibility consists of reducing the dimension (and thus the memory footprint) of the forward operator $\imop$ with $\cl P$ random Gaussian projections and computes $\bs y := \bsmat{A}\bs{v}$ and $\bar{\bs y}:=\bsmat A\bar{\bs v}$, which give
\begin{align} 
    \bs y &= \bsmat{A} \acqop(\cl X) + \bsmat{A} \bs\xi, \\
    \bar{\bs y} &= \bsmat A \imop(\bs\sigma) + \bsmat A \bar{\bs\xi}, \label{eq:gaussian_im}
\end{align}
with $\bs y$, $\bar{\bs y} \in \Cbb^{\cl P}$ and $\bsmat A \in \Cbb^{\cl P \times VB}$ with $A_{jk} \sim_\iid \cl{CN} (0,1/\cl P)$ \citep{kartik17}. 
Simple derivations show that, with visibility weighting and in expectation with respect to the coefficients of $\bsmat A$, the noise statistics are $\bsmat{AW}\bar{\bs\xi} \sim \cl{CN}(\bs 0, \tau^2 \cl P^{-1} \norm{\bs w}{2}^2 \Id)$. In conclusion, whichever the initial visibility weighting choice, natural or uniform, only a global rescaling of the original weights is needed to ensure unit variance of the noise on the final measurements. Under this normalisation constraint, natural visibility weighting follows as $\bs w_{\rm n} = \tau^{-1} \sqrt{\cl P/VB}\,\bs 1_{VB}$, while uniform visibility weighting reads $\bs w_{\rm u} = \tau^{-1} \sqrt{\cl P}\, \bs\rho^{-1/2}/\norm{\bs\rho^{-1/2}}{2}$.

Unfortunately, as will be explained in Section~\ref{sec:complexity}, $\bsmat A$ is dense and makes its application too computationally expensive in practice. 
This random Gaussian compression of the classical scheme is illustrated on the right-side of Fig.~\ref{fig:acquisition}(a). 

\subsection{Antenna-based MROP} \label{sec:mrop1}

This section begins by introducing the antenna-based MROP model outlined in \citet{leblanc2024b}. It then deepens the analysis by exploring the interplay between visibility weighting and random beamforming and examining the noise statistics associated with MROPs.

\subsubsection{Revisiting the Sensing Model}
\label{sec:revisiting}

The compressive sensing scheme, namely \emph{modulated rank-one projections} (MROP), proposed in \citet{leblanc2024b} differs from the post-correlation compression schemes described above because the compression is applied at the antenna measurements level. 
It consists of combining two compression layers: \emph{(i)} \emph{random beamforming} trading the $V$-dependence of the data dimension in \eqref{eq:classic_acq} for a smaller number $P$ of ROPs, and \emph{(ii)} \emph{random modulations} trading the number of batches $B$ for $M$ modulations of the ROP vectors obtained for each batch. 

\paragraph*{Layer 1: random beamforming.}

Let us consider a \emph{phased-array} model with two sets of random sketches of the measurement vector $\bs\mu(t):=\bc A^*\bs x(t)$ and $\bs\nu(t):=\bc B^*\bs x(t)$ using the matrices $\bc A:=(\bs\alpha_1,\ldots,\bs\alpha_P)$ and $\bc B:=(\bs\beta_1,\ldots,\bs\beta_P)$ where, $\forall p \in\upto{P}$, $\bs\alpha_p = (\alpha_{p1},\ldots,\alpha_{pQ})^\top \sim_\iid \bs u$ for some random vector $\bs u \in \bb C^Q$ applying phase delays with $u_q \sim_iid P^{-1/4} \exp(\im \cl U[0,2\pi))$, \ie $\bb E[u_q]=0$ and $|u_q|^2=1/\sqrt{P}$, $\forall q\in\upto{Q}$ (and similarly for $\bs\beta_p$), and more specifically their sampling $\bs\mu_b[i]$ and $\bs\nu_b[i]$ obtained in the same way as $\bs x_b[i]$ was obtained in \eqref{eq:xb}. Aggregating their elementwise products in time gives
\begin{equation} \label{eq:proj_prod}
\bs y_b' := \frac{1}{I} \sum_{i=1}^{I} \bs\mu_b[i] \odot \bs\nu_b^*[i],
\end{equation}
whose components write 
\begin{equation} \label{eq:31_sample_rop}
    y_{pb}' = \frac{1}{I} \sum_{i=1}^{I} \bs\alpha_p^* \bs x_b[i] \bs x_b^*[i] \bs\beta_p = \bs\alpha_p^* \bsmat C_b \bs\beta_p,~~\forall p \in \upto{P}.
\end{equation}  
In \eqref{eq:31_sample_rop}, $\bs\alpha_p^* \sampcovmat_b \bs\beta_p := \langle \bs\alpha_p\bs\beta_p^*, \sampcovmat_b \rangle_{\rm F}$ is called a \emph{rank-one projection} (ROP) of the sample covariance matrix $\bs C_b$ defined in \eqref{eq:31_empi_covmat} because it amounts to projecting it onto the rank-one matrix $\bs\alpha_p\bs\beta_p^*$ \citep{cai2015rop,chen2015exact}.
As clear from \eqref{eq:proj_prod}, the computational complexity of the ROP computation scales with the number of antennas $Q$.
This characteristic sets the antenna-based MROP model apart from the visibility-based MROP model discussed in Section~\ref{sec:mrop2}.

Analogously to the upper triangular selection made for the classical model in Section~\ref{sec:classic}, a \emph{debiasing} step is necessary. In the ROP measurements of \eqref{eq:31_sample_rop}, debiasing consists of removing the dependency in the diagonal $\sampcovmat_b^\rmd := \diag(\diag \sampcovmat_b)$ including $Q$ repetitions of the DC component $\cl F[\sigma^\circ](\bs 0)$ and the diagonal covariance matrix of the noise $\noisemat$.
Both quantities can be estimated from a limited set of additional measurements \citep{leblanc2024b}.
Gathering the $B$ measurement vectors $(\bs y_b)\btoB$ with each $\bs y_b := (y_{pb})_{p=1}^P \in \bb C^P$ containing the $P$ debiased ROPs yields the \emph{random beamforming} acquisition model
\begin{equation} \label{eq:acqopn}
 \bs y = (\bs y_1^\top, \ldots, \bs y_B^\top)^\top := \acqopn(\cl X) + \wt{\bs\xi} \in \bb C^{PB},
\end{equation}
where $\wt{\bs\xi}$ is additive noise and with 
\begin{equation}
y_{pb} := y_{pb}' - \bs\alpha_{pb}^* \sampcovmat_b^\rmd(\cl X_b) \bs\beta_{pb} + \wt\xi_{pb},\ \forall p \in \upto{P},\ \forall b \in \upto{B}.
\end{equation}

The imaging model associated with \eqref{eq:acqopn} corresponds to a linear transform of the visibility vector and writes 
\begin{equation} \label{eq:32_yb}
    \bar{\bs y}_b = \bsmat R_b \bar{\bs v} = \bsmat R_b \bsmat G_b \bsmat{FZ}\bs\sigma + \bsmat R_b \bar{\bs\xi}_b,
\end{equation}
where each row of the matrix $\bsmat R_b$ computes a single ROP of the (debiased) \emph{hollow} matrix $\sampcovmat_b^\rmh := \sampcovmat_b-\sampcovmat_b^\rmd$. Applied to the upper triangular matrix $\bc T \sampcovmat_b$ that is vectorised in $\bsmat G_b \bsmat{FZ}\bs\sigma$ for the classical model described in Section~\ref{sec:classic},  where $\bsmat G_b \in \Cbb^{V \times 4N}$ is the convolutional interpolation operator corresponding only to batch $b$, 
$\bsmat R_b: \bs u \mapsto \bsmat R_b \bs u$ can be defined as 
\begin{equation} \label{eq:Rb}
\resizebox{.9\linewidth}{!}{$
    (\bsmat R_b \bs u)_p = \vect(\bc T\bs\alpha_{pb} \bs\beta_{pb}^*)^\top \bs u + (\vect(\bc T\bs\beta_{pb} \bs\alpha_{pb}^*)^\top \bs u )^*,~~\forall p \in \upto{P}.
$}
\end{equation}
The $B$ ROPed batches can then be concatenated as 
\begin{equation} \label{eq:sep_rop1}
    \bar{\bs y} = \begin{bmatrix}
        \bar{\bs y}_1 \\ \vdots \\ \bar{\bs y}_B
    \end{bmatrix} = \begin{bmatrix}
        \bsmat R_1 & & \\ 
        & \ddots & \\ 
        & & \bsmat R_b
    \end{bmatrix} \left( \begin{bmatrix}
        \bsmat G_1 \\ \vdots \\ \bsmat G_B
    \end{bmatrix} \bsmat{FZ}\bs{\sigma} + \begin{bmatrix}
        \bar{\bs\xi}_1 \\ \vdots \\ \bar{\bs\xi}_B
    \end{bmatrix} \right) = \bsmat{DGFZ}\bs\sigma + \bsmat D \bar{\bs\xi},
\end{equation}
or, equivalently,
\begin{equation} \label{eq:sep_rop2}
    \bar{\bs y} = \wt\imop(\bs\sigma) + \bsmat D \bar{\bs\xi},
\end{equation}
with $\bar{\bs y} \in \Cbb^{PB}$, and $\wt\imop:=\bsmat D \imop$. 
The imaging model \eqref{eq:sep_rop2}, coined \emph{concatenated} ROP (CROP), consists of adding the CROP operator $\bsmat D \in \Cbb^{PB \times VB}$ to the classical model \eqref{eq:classic} sensing the visibilities.
The impact of the CROP operator $\bsmat D$ on the noise statistics is less trivial than in the previously discussed models and is therefore thoroughly analysed in Section~\ref{sec:mrop_noise}.

CROP can be considered a compression model of its own. However, one expects a significant redundancy in the information carried by the $B$ batches of $P$ projections, as these arise from non-independent Fourier data of the image to be formed.\footnote{This can be evidenced by the low-rankness of the $V \times B$ matrix resulting from concatenating vectorised versions of the $B$ covariance matrices. A simple singular value decomposition on such a matrix in the simulation setup of Sections \ref{sec:complexity}-\ref{sec:setting} confirms this low-rankness assumption.}
This leaves the opportunity for additional compression and has motivated the introduction of the MROP model.

\paragraph*{Layer 2: random modulations.}

MROP follows from introducing a second layer of dimensionality reduction, consisting in applying $M$ random modulations of the ROP vectors before integrating them across batches, \ie we compute 
\begin{equation} \label{eq:im_mrop1}
    \bar{\bs z}_m := \sum\btoB \gamma_{mb} \bar{\bs y}_b,
\end{equation} 
with the \emph{modulation vectors} $\bs\gamma_m \sim_\iid \bs\gamma$, $\forall m \in \upto{M}$, each component $\gamma_b$ of $\bs\gamma$ being \iid as the (normalised) Bernoulli distribution $M^{-1/2} \cl U\{\pm 1\}$. 
From the CROP model \eqref{eq:sep_rop2}, this can also be viewed as
\begin{equation*}
    \bar{\bs z}_m = \big[\gamma_{m1} \Id, \ldots, \gamma_{mB} \Id \big]\,\bar{\bs y} = (\bs\gamma_m^\top \otimes \Id_P) \bar{\bs y}.
\end{equation*}
Moreover, the $M$ modulations can be concatenated as 
\begin{equation} \label{eq:Gamma}
    \bar{\bs z} = \begin{bmatrix}
        \bar{\bs z}_1 \\ \vdots \\ \bar{\bs z}_M
    \end{bmatrix} = (\bs\Gamma^\top \otimes \Id_P) \bar{\bs y} = \bsmat M \bar{\bs y},
\end{equation} 
with $\bs\Gamma := [\bs\gamma_1, \ldots, \bs\gamma_M] \in \{ \pm 1 \}^{B\times M}$, $\bsmat M := \bs\Gamma^\top \otimes \Id_P \in \{ \pm 1,0 \}^{PM \times PB}$, and $\bar{\bs z} \in \Cbb^{PM}$. 
The resulting acquisition and imaging models respectively write 
\begin{align} 
    \bs z &:= \bsmat M \acqopn(\cl X) + \bsmat M \wt{\bs\xi}, \label{eq:acq_mrop} \\
    \bar{\bs z} &= \bsmat M \wt\imop(\bs\sigma) + \bsmat{MD} \bar{\bs\xi}, \label{eq:im_mrop}
\end{align} 
which turn into simply applying the \emph{modulation operator} $\bsmat M$ to the random beamforming acquisition model \eqref{eq:acqopn} and to the CROP model \eqref{eq:sep_rop2}. 
The imaging model \eqref{eq:im_mrop} is illustrated in Fig.~\ref{fig:acquisition}(d).
The computations of this random beamforming acquisition of MROP \eqref{eq:acq_mrop} are illustrated in Fig.~\ref{fig:acquisition}(b).
The notion of \emph{batchwise} computation, which is critical to memory efficient implementations and is illustrated in Fig.~\ref{fig:acquisition}, is explained in Section~\ref{sec:remarks}.
Starting from $\bsmat Z^{(0)} = \bsmat 0_{P \times M}$, the random modulations are summed batchwise as $\bsmat Z^{(b)} = \bsmat Z^{(b-1)} + \bs y_b \times \bs\gamma_b$ to remain at a fixed data dimension $PM$. The MROP vector $\bs z$ is the vectorisation of the final matrix $\bsmat Z^{(B)}$.

\paragraph*{IROP: the asymptotic case.} \label{sec:irop} 

We also present the asymptotic case where no modulation is applied before time integration, \ie $\gamma_{mb} = 1$ for all $b$ and a single $m$ in \eqref{eq:im_mrop1}, so that $\bs z = \sum\btoB \bs y_b \in \Cbb^P$. This model is coined \emph{integrated} ROP (IROP). It amounts to replacing $\bs\Gamma$ with $\bs 1_B$ in \eqref{eq:Gamma}, yielding $\bsmat M = \bs 1_B^\top \otimes \Id_P = [\Id, \ldots, \Id]$, hence writing 
\begin{align}
    \bs z &= \big[\Id, \ldots, \Id \big]\, \acqopn (\cl X) + \big[\Id, \ldots, \Id \big]\, \wt{\bs\xi} \\
    \bar{\bs z} &= \bsmat R \imop(\bs\sigma) + \bsmat R \bar{\bs\xi}, \label{eq:irop}
\end{align}
with $\bsmat R := \big[\Id, \ldots, \Id \big]\,\bsmat D = \big[\bsmat R_1, \ldots, \bsmat R_b\big] \in \Cbb^{P \times VB}$. 

In \citet{leblanc2024b}, we theoretically demonstrated that the total forward operator $\bsmat R \imop$ satisfies a variant of the \emph{restricted isometry property} (RIP) \citep{chen2015exact}, the $\RIPto$, a property ensuring the recovery of sparse images from their compressive observations. This method was shown to achieve reduced sample complexities compared to the classical uncompressed scheme of \eqref{eq:classic}. However, as this asymptotic IROP case brings the same number of measurements as an MROP approach with $M=1$, one \emph{a priori} expects that the number $P$ of IROPs necessary for accurate image estimation is larger than in an MROP approach with $M>1$. A more detailed discussion of relative model complexities is deferred to Section \ref{sec:complexity}, which will reveal the advantage of MROP.

\subsubsection{Compatibility with Visibility Weighting} \label{sec:mrop_weighting}

This section extends the antenna-based MROP acquisition model by showing that visibility weighting is still compatible with random beamforming but at the price of an increased acquisition cost. \\

One issue with the antenna-based acquisition of MROPs using random beamforming, as clear from the definition of the random beamforming acquisition operator $\acqopn$ in \eqref{eq:acqopn} and illustrated in Fig.~\ref{fig:acquisition}(b), is that there is no access to the visibility matrices $\{\sampcovmat_b^\rmh \}\btoB$. 
The visibilities thus seemingly cannot be weighted before being compressed. 
However, the weighting of the visibility matrix $\sampcovmat_b^\rmh$ of batch $b$ can be distributed within the sketching vectors by decomposing the weighting matrix $\wt{\bsmat W}_b$ (introduced in Section~\ref{sec:weighting}) as a sum of rank-one matrices, \ie $\wt{\bsmat W}_b = \sum_{r=1}^R \wt{\bs w}_{br} \wt{\bs w}_{br}^*$.
Indeed, the $pb$-th \emph{visibility-weighted} CROP measurement obtained from the random beamforming acquisition can be written as 
\begin{align} \label{eq:weighting}
\begin{split}
    \bar y_{pb,\wt{\bsmat W}_b} &= \bs\alpha_{pb}^* (\wt{\bsmat W}_b \odot \sampcovmat_b^\rmh) \bs\beta_{pb} \\
&= \bs\alpha_{pb}^* (\sum_{r=1}^R \wt{\bs w}_{br} \wt{\bs w}_{br}^* \odot \sampcovmat_b^\rmh) \bs\beta_{pb} \\
    &= \sum_{r=1}^R (\bs\alpha_{bp} \odot \wt{\bs w}_{br})^* \sampcovmat_b^\rmh (\bs\beta_{pb} \odot \wt{\bs w}_{br}).
\end{split}
\end{align}
where the operation $\odot$ is an elementwise multiplication.

Eq.~\eqref{eq:weighting} shows that visibility weighting comes at the price of multiplying the initial number of ROPs by the rank $R$ of the weighting matrix.
This suggests that a trade-off between \emph{(i)} increasing $R$ for an improved image recovery by visibility weighting, and \emph{(ii)} decreasing $R$ to decrease the acquisition cost, discussed in Section~\ref{sec:complexity}, must be found in order to still benefit from a compressive imaging scheme. 
However, this increased acquisition cost does not impact the memory requirements or the image reconstruction process.

A drawback that the MROP model shares with the dirty image and random Gaussian compression models is that the visibility weighting is fixed forever once the weighted MROPs have been acquired.

\subsubsection[Noise Statistics on MROPs]{Noise Statistics on MROPs} \label{sec:mrop_noise}

The noise statistics play a crucial role in image recovery, as detailed in Section~\ref{sec:setting}. 

\begin{landscape}
\begin{figure}
\centering
\includegraphics[width=\linewidth]{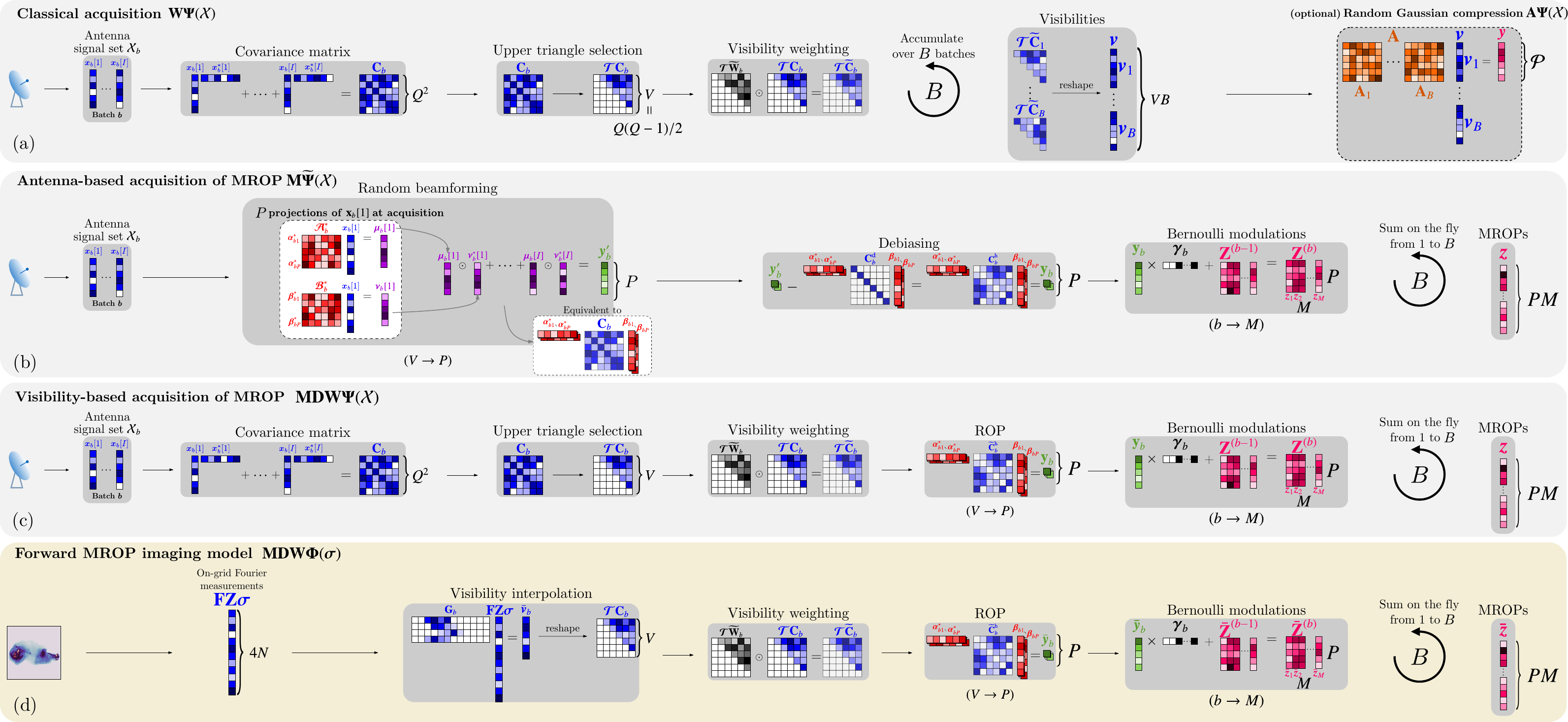}
\put(-662,249){\small Section~\ref{sec:classic}}
\put(-95,249){\small Section~\ref{sec:post_corr}}
\put(-662,154){\small Section~\ref{sec:mrop1}}
\put(-662,84){\small Section~\ref{sec:mrop2}}
\caption{Computation schemes with batchwise implementations. (a-c) computations at the acquisition: from the antenna signals $\cl X$ to the (compressed) visibilities. 
(a) \textbf{Classical acquisition}. For each batch $b$, the sample covariance matrix is computed as $\sampcovmat_b = \tinv I \sum_{i=1}^I \bs x_b[i] \bs x_b[i]^*$. 
Selecting the upper triangular part of the covariance matrices provides the visibilities in $\bc T \sampcovmat_b$. 
The visibilities can be weighted as $\bc T\wt{\bsmat W}_b \odot \bc T \sampcovmat_b = \bc T \wt{\sampcovmat}_b$. 
An \emph{optional} \textbf{random Gaussian compression} of the (vectorised) visibilities $\bs v$ is possible using an \iid random Gaussian matrix $\bsmat A$ as $\bs y =\bsmat{A}\bs v$.
Its batchwise implementation (not shown here) implies aggregating the projections over batches as $\bs{y}=\sum\btoB \bsmat{A}_b \bs{v}_b$. 
(b) \textbf{Antenna-based acquisition of MROP}. 
For each batch $b$, $P$ (biased) ROPs of the covariance matrix are obtained as $y_{pb}' = \tinv I \sum_{i=1}^I \bs\alpha_{pb}^* \bs x_b[i] \bs x_b[i]^* \bs\beta_{pb}$, $\forall p\in\upto{P}$. 
The ROPs are debiased as $y_{pb} = y_{pb}'-\bs\alpha_p^* \sampcovmat_b^\rmd \bs\beta_p$. 
$M$ copies of the ROP vector $\bs y_b$, with random signs encoded into $\bs\gamma_b = (\gamma_{mb})_{m=1}^M$ and represented by the black and white cells, are opened as $\bs y_b\times \bs\gamma_b$.
These $PM$ MROPs are progressively summed over the $B$ batches as $\bsmat Z^{(b)} = \bsmat Z^{(b-1)} + \bs y_b \times \bs\gamma_b$ with $\bsmat Z^{(0)}=\bsmat 0_{P\times M}$. 
There is no visibility weighting here. 
(c) \textbf{Visibility-based acquisition of MROP}.
For each batch $b$, the sample covariance matrix is computed like in (a), giving direct access to the upper triangular covariance matrix $\bc T\sampcovmat_b$ and allowing for a weighting as $\bc T\wt{\bsmat W}_b \odot \bc T\sampcovmat_b = \bc T\wt\sampcovmat_b$. 
Then, symmetrising the weighted visibility matrix as $\bc T\wt\sampcovmat_b + (\bc T\wt\sampcovmat_b)^*=\wt\sampcovmat_b^\rmh$, $P$ random ROPs of the weighted interferometric matrix are obtained as $y_{pb} = \bs\alpha_{pb}^* \wt\sampcovmat_b^\rmh \bs\beta_{pb}$, $\forall p\in\upto{P}$. 
From that, the random modulations are the same as in (b). 
(d) \textbf{Forward MROP imaging model}. 
Starting from the FFT of the (zero-padded) discrete image $\bsmat{FZ}\bs{\sigma}$, the visibilities related to the batch $b$ are interpolated as $\bar{\bs v}_b = \bsmat G_b \bsmat{FZ}\bs{\sigma}$ then reshaped to get the upper triangular matrix $\bc T \sampcovmat_b$. 
The next operations are the same as in (c) with the weighting, ROPs, and random modulations.
A key observation in Fig.~\ref{fig:acquisition}(b-d) is that the number of measurements never exceeds $PM$ both during the acquisition and the imaging because the ROPs and modulations can be computed and integrated batchwise.
}
\label{fig:acquisition}
\end{figure}
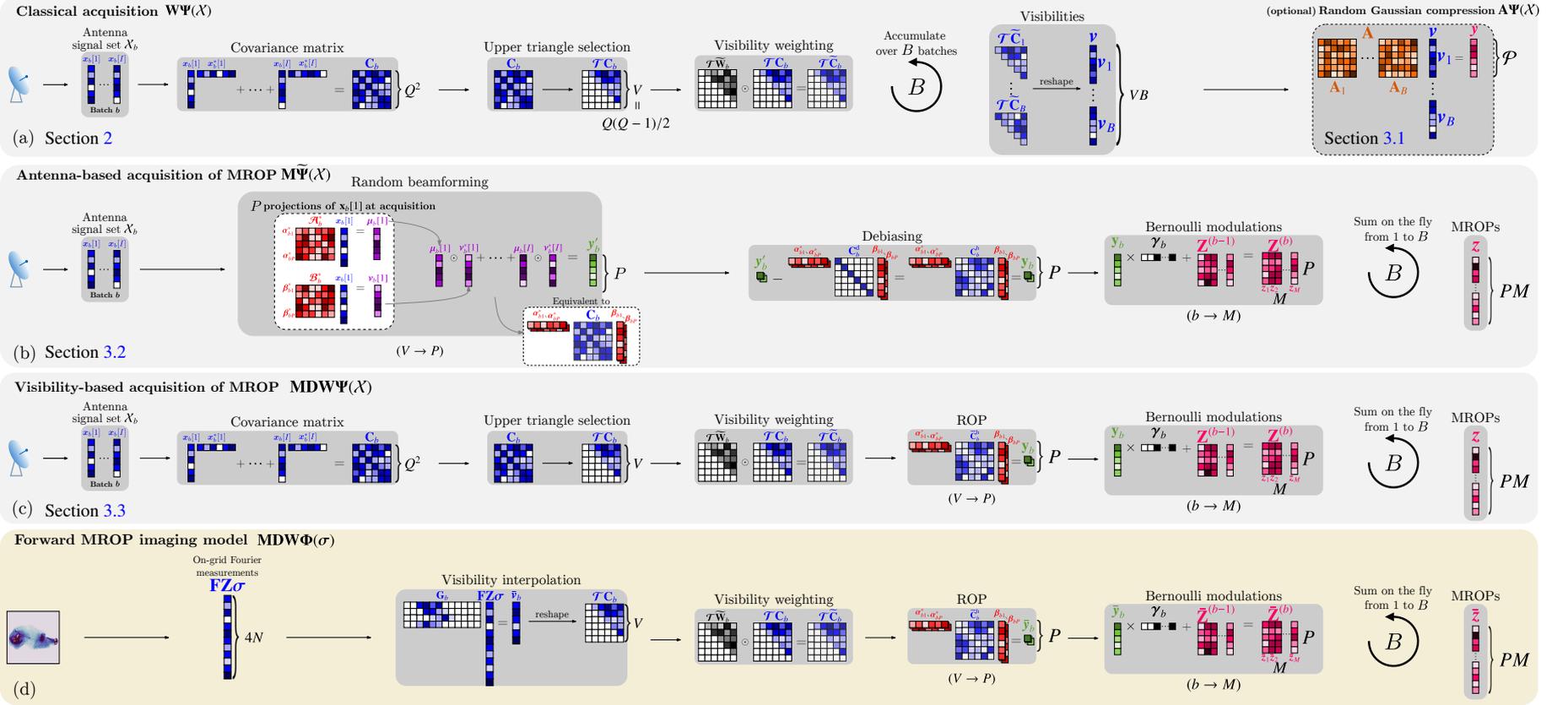
\end{landscape}

Now that we have shown the MROP model is compatible with visibility weighting, this section deepens the understanding of the MROP model by deriving the noise statistics on visibility-weighted MROPs. \\

As a reminder from Section~\ref{sec:weighting}, the noise on the weighted visibilities is $\bsmat W\bar{\bs\xi} \sim \cl{CN} (\bs 0, \tau^2 \diag(\bs w^2))$.
Let us divide the next derivations into two steps by first studying the noise statistics in the CROP model $\bar{\bs\xi}' := \bsmat{DW}\bar{\bs{\xi}} \in \Cbb^{PB}$, and then in the MROP model $\bar{\bs\xi}'':=\bsmat M \bar{\bs\xi}' = \bsmat{MDW}\bar{\bs\xi} \in \Cbb^{PM}$. 
First, we study the impact of the CROP operator $\bsmat D$ on the weighted noise statistics.
By linearity, $\bb E [\bar{\bs\xi}'] = \bsmat{DW} \bb E [\bar{\bs\xi}] = \bs 0$.
From the definition of $\bsmat D$, and making an abuse of notation by writing $\bar\xi'_{pb} = \xi'_{p+bP}$ for brevity, the $pb$-th element of $\bar{\bs\xi}'$ writes as
\begin{equation}
    \bar\xi'_{pb} = (\bsmat R_b \bsmat W_b \bar{\bs\xi}_b)_p = \bs\alpha_{pb}^* \bs\Xi_b \bs\beta_{pb} = \sum_{i,j=1, i\neq j}^Q \alpha_{pbi}^* \bar\Xi_{bij} \beta_{pbj},
\end{equation}
where $\bsmat W_b$ is the $b$-th block of $\bsmat W$,  $\bar{\bs\Xi}_b$ is the hermitian matrix
containing the noise vector $\bsmat W_b \bar{\bs\xi}_b$ in its upper triangular part plus its conjugate in the lower triangular part.
The element of the covariance  matrix of $\bar{\bs\xi}'$ associated with the indices $p,b$ and $p',b'$ writes as
\begin{align} \label{eq:ab_fixed}
\begin{split}
    \E[\bar{\bs\xi}]{\bar\xi'_{pb} \bar\xi_{p'b'}^{'*}}
    &= \sum_{i,j,k,l=1, i\neq j, k\neq l}^Q \alpha_{pbi}^* \beta_{pbj} \alpha_{p'b'k} \beta_{p'b'l}^* \E[\bar{\bs\xi}]{\bar\Xi_{bij} \bar\Xi_{b'kl}^*} \\
    &= \sum_{i,j=1, i\neq j}^Q \alpha_{pbi}^* \beta_{pbj} \alpha_{p'bi} \beta_{p'bj}^* \delta_{bb'} \tau^2 \wt{W}_{bij}^2.
\end{split}
\end{align}
Taking the expectation with respect to $\alpha$ and $\beta$ in \eqref{eq:ab_fixed} yields
\begin{align} \label{eq:ab_random}
\begin{split}
    \E[\alpha,\beta,\bar{\bs\xi}]{\bar\xi'_{pb} \bar\xi_{p'b'}^{'*}}
    &= \tau^2 |\alpha|^2 |\beta|^2 \delta_{bb'} \delta_{pp'} \norm{\wt{\bsmat W}_b}{\rm F}^2
\end{split}
\end{align}
and, after injecting $|\alpha|^2=|\beta|^2=P^{-1/2}$, demonstrates that 
\begin{equation}
\E[\alpha,\beta,\bar{\bs\xi}]{\bar{\bs\xi}' \bar{\bs\xi}^{'*}} = \frac{\tau^2}{P} 
\begin{bmatrix} \norm{\wt{\bsmat W}_1}{\rm F}^2 \Id_Q & & \\
& \ddots & \\
& & \norm{\wt{\bsmat W}_B}{\rm F}^2 \Id_Q
\end{bmatrix}.
\end{equation}
Defining $\bs w_b = \vect(\bc T \wt{\bsmat W}_b)$, so that $\norm{\wt{\bsmat W}_b}{\rm F}^2 =2\norm{\bs w_b}{2}^2$, the covariance matrix of the noise $\bsmat{R}_b\bsmat{W}_b\bar{\bs\xi}_b$ is thus, in expectation over $\alpha$ and $\beta$, the identity matrix up to a normalisation $2\tau^2 P^{-1} \norm{\bs w_b}{2}^2$.
In conclusion, whichever the initial visibility weighting choice, natural or uniform, only a global rescaling per batch of the original weights is needed to ensure unit variance of the noise on the final measurements. Under this simple normalisation constraint, natural visibility weighting thus follows as $\bs w_{\rm n} = \tau^{-1} \sqrt{P/2V} \,\bs 1_{VB}$, while uniform visibility weighting reads, for each batch $b\in\upto{B}$, $\bs w_{{\rm u},b} = \tau^{-1} \sqrt{P/2} \bs\rho_b^{-1/2}/\norm{\bs\rho_b^{-1/2}}{2}$, where $\bs\rho_b \in \Rbb^V$ is the restriction to batch $b$ of the vector of density $\rho$.

Secondly, let us study the impact of the modulation operator $\bsmat M$ on the noise statistics.
By linearity, $\bb E [\bar{\bs\xi}''] = \bsmat M \bb E [\bar{\bs\xi}'] = \bs 0$.
From the definition of $\bsmat M$, the $pm$-th noise element in the MROP model writes as
\begin{equation}
    \bar\xi''_{pm} = \sum\btoB \gamma_{mb} \bar\xi_{pb}'.
\end{equation}

Re-using \eqref{eq:ab_random}, the element of the covariance matrix of $\bar{\bs\xi}''$ associated with the indices $p,m$ and $p',m'$ writes as
\begin{align} \label{eq:gamma_fixed}
\begin{split}
    \E[\alpha,\beta,\bar{\bs\xi}]{\bar\xi''_{pm} \bar\xi_{p'm'}^{''*}}
    &= \sum_{b,b'=1}^B \gamma_{mb} \gamma_{m'b'} \E[\alpha,\beta,\bar{\bs\xi}]{\bar\xi'_{pb} \bar\xi_{p'b'}^{'*}} \\
    &= \sum_{b,b'=1}^B \gamma_{mb} \gamma_{m'b'} \frac{\tau^2}{P} \delta_{bb'} \delta_{pp'} \norm{\wt{\bsmat W}_b}{\rm F}^2 \\
    &= \frac{\tau^2}{P} \delta_{pp'} \sum_{b=1}^B \gamma_{mb} \gamma_{m'b} \norm{\wt{\bsmat W}_b}{\rm F}^2. 
\end{split}
\end{align}
With $\sum\btoB \norm{\wt{\bsmat W}_b}{\rm F}^2 = 2 \norm{\bs w}{2}^2$ and $|\gamma|^2=M^{-1}$, taking the expectation with respect to $\gamma$ in \eqref{eq:gamma_fixed} demonstrates that $\E[\alpha,\beta,\gamma,\bar{\bs\xi}]{\bar{\bs\xi}'' \bar{\bs\xi}''^*} = 2 \tau^2 P^{-1} M^{-1} \norm{\bs w}{2}^2 \Id_{PM}$. 
Normalising the weights by ensuring $\norm{\bs w}{2}^2 = \tau^{-2} PM/2$ also ensures a unit standard deviation.
In conclusion, whichever the initial visibility weighting choice, natural or uniform, only a global rescaling of the original weights is needed to ensure unit variance of the noise on the final measurements. Under this simple normalisation constraint, natural visibility weighting follows as $\bs w_{\rm n} = \tau^{-1} \sqrt{PM/2VB} \,\bs 1_{VB}$, while uniform visibility weighting reads $\bs w_{\rm u} = \tau^{-1} \sqrt{PM/2}\,\bs\rho^{-1/2}/\norm{\bs\rho^{-1/2}}{2}$. We finally note that the IROP case follows by taking $M=1$ in the above relations.

\subsection{Visibility-based MROP: Back to Post-Correlation} 
 \label{sec:mrop2}

We introduce an alternative acquisition method for MROP and its CROP and IROP variants, based on visibilities rather than antenna-based random beamforming (see Section~\ref{sec:mrop1}), as shown in Fig.~\ref{fig:acquisition}(c). 
This approach is of the same nature as the models discussed in Section~\ref{sec:post_corr} and remains compatible with existing correlators \citep{Thompson17}. 
Within each batch $b$, the visibilities $\bc T\sampcovmat_b$ are weighted from an elementwise multiplication with an upper triangular weighting matrix $\bc T \wt{\bsmat W}_b$ as $\bc T \wt{\sampcovmat}_b = \bc T \wt{\bsmat W}_b \odot \bc T\sampcovmat_b$, like in the classical model \eqref{eq:classic_acq}.
Next, the Hermitian matrix $\wt{\sampcovmat}_b^\rmh = \bc T \wt\sampcovmat_b + (\bc T\wt\sampcovmat_b)^*$, containing the weighted visibilities and their complex conjugates, is projected onto $P$ rank-one matrices $\{ \bs\alpha_{pb}\bs\beta_{pb}^* \}_{p=1}^P$, yielding the ROP vector $\bs y_b = \{ \bs\alpha_{pb}^* \wt{\sampcovmat}_b^\rmh \bs\beta_{pb} \}_{p=1}^P$.
This ROP vector is distributed over $M$ channels with random modulations and integrated with the previous ones as $\bsmat Z^{(b)} = \bsmat Z^{(b-1)} + \bs y_b \times \bs\gamma_b$ starting from $\bsmat Z^{(0)} = \bs 0_{P \times M}$ for $b \in \upto{1,B}$, and where $\bs\gamma_b := (\gamma_{mb})_{m=1}^M$ are random signs. 
The final vector $\bs z \in \Cbb^{PM}$ follows the acquisition model
\begin{equation}
    \bs z = \bsmat{MDW}\acqop(\cl X),
\end{equation}
while the imaging model remains the same as in \eqref{eq:im_mrop}.

This method solves the problem of applying visibility weighting to MROPs, which was the case of the antenna-based MROP method as discussed in Sec.~\ref{sec:mrop_weighting}. 
As for the antenna-based acquisition of MROPs, the weighting is fixed forever once the weighted MROPs have been acquired.
\section{Computational Costs and Memory Requirements} \label{sec:complexity}

\begin{figure*}
    \centering
    \includegraphics[width=0.9\linewidth]{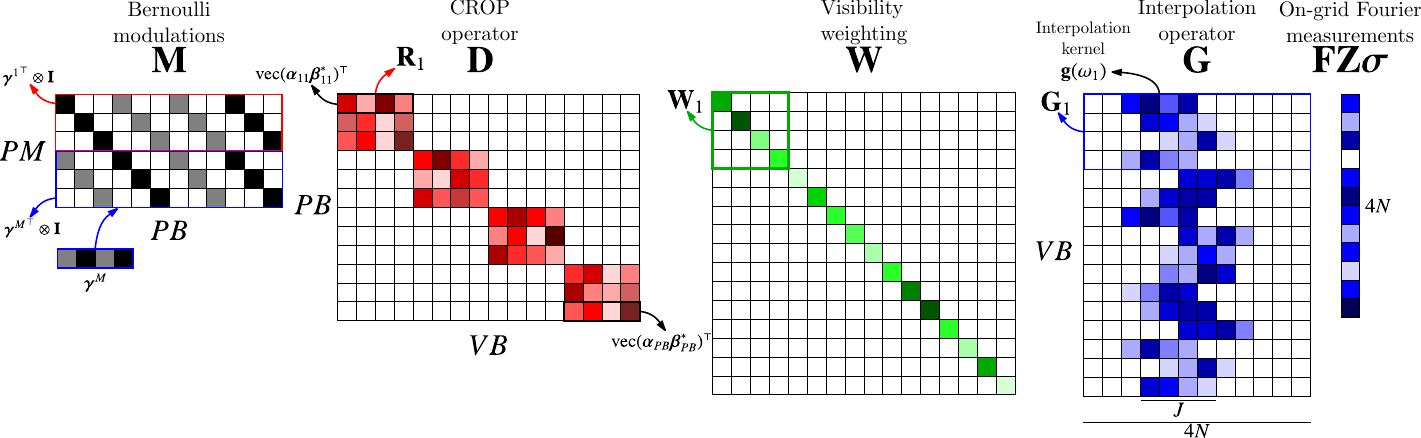}
    \caption[Structure of the MROP operators.]{Structure of the operators in the MROP imaging model. An abuse of notation is made to simplify the illustration by considering that the interpolation operator $\bsmat G$ directly outputs the hermitian visibility matrix $\sampcovmat_b^\rmh$ instead of the upper triangular matrix $\bc T \sampcovmat_b$, \ie $\bsmat G_b \bsmat{F}\bsmat{Z}\bs{\sigma}=\vect(\sampcovmat_b^\rmh)$.
    The $VB$ visibilities are obtained by interpolating with $\bsmat G$, and at a cost $\cl O(JVB)$, between the on-grid frequencies of the image obtained from a (real) FFT of the zero-padded image $\bsmat{F}\bsmat{Z}\bs{\sigma}$ at a cost $\cl O(2N\log N)$. Each visibility is assigned a weight using the diagonal visibility weighting operator $\bsmat W \in \Rbb^{VB\times VB}$ at a cost $\cl O(VB)$. The simplification of this illustration appears in the CROP operator $\bsmat D$ whose blocks $\{ \bsmat R_b \}\btoB$ have here rows that are the vectorisation of a rank-one matrix, \ie $(\bsmat R_b)_p = \vect(\bs\alpha_{pb} \bs\beta_{pb}^*)^\top$ instead of the more complicated expression in \eqref{eq:Rb}. The computation of the CROP operator $\bsmat D$ requires $\cl O(2PVB)$ operations. The MROPs are finally obtained by applying the modulation operator $\bsmat M$ at a negligible cost $\cl O(PMB)$.}
    \label{fig:operators}
\end{figure*}

\emph{Note: The reader more interested in experimental validation is invited to skip directly to Sections~\ref{sec:setting}-\ref{sec:results}.}\\

\noindent This section analyses the computational costs and memory requirements 
for all the models introduced in Sections \ref{sec:classic}-\ref{sec:compression} at both acquisition (see Table \ref{tab:acquisition}) and image reconstruction (see Table~\ref{tab:imaging}). 
The analysis is supported by Fig.~\ref{fig:acquisition} and Fig.~\ref{fig:operators} which aim to illustrate the involved computations.
Firstly, preliminary remarks in Section~\ref{sec:remarks} explain how combining batchwise, precomputation-free, and on-the-fly computation approaches ensure minimal memory requirements. 
Secondly, Section~\ref{sec:tab1} describes the example scenario considered in the simulations of Sections~\ref{sec:setting}-\ref{sec:results} to serve as a numerical illustration accompanying the complexity analysis that follows in Sections~\ref{sec:comp_classic}-\ref{sec:comp_mrop}. 
Section \ref{sec:firstconclusion} draws first conclusions, with models such as the random Gaussian projections and IROP deemed non-viable due to their expected high computational cost, and models such MROP and its CROP variant presenting a high potential for data dimensionality reduction and reduced memory requirements over the classical and BDA models. 
Finally, Section~\ref{sec:dirty_remark} concludes with remarks on blessing and curse of the dirty image model, explaining why it is not retained for our numerical experiments.

We note that throughout, computational costs are computed in units of the cost of a complex addition or multiplication. In other words, it is reported as to the number of such operations. 
Memory requirements are counted in units of the memory required to store a complex-valued number (\emph{e.g.}~16 bytes in double precision). 

\subsection{Memory Efficient Computations} \label{sec:remarks}

While minimising the final number of measurements $D$ serves the memory requirements for the long-term storage, limiting the memory requirement during acquisition and image reconstruction is equally important. 
This section describes the concepts of \emph{batchwise} and \emph{precomputation-free} implementations, both leveraging \emph{on-the-fly} computation to reduce memory requirements.

\noindent\textbf{Batchwise computation at acquisition.}
Firstly, for any of the post-correlation compression models described in Section~\ref{sec:compression}, including the visibility-based MROP and variants, a naive acquisition approach consists in accumulating all the $VB$ visibilities first, and then applying compression. 
This however implies memory requirements $\cl O(VB)$, partly defeating the purpose of the compression. 
A simple solution is to adopt a batchwise implementation, consisting in compressing the covariance matrix of each batch successively and summing the compressed data on the fly over batches, hence limiting the memory requirement to  $\cl O(\max(V,\alpha D))$ with $\alpha=1$ and $\alpha=1/2$ for models with complex- and real-valued data respectively. 
Fig.~\ref{fig:acquisition}(c) illustrates the batchwise implementation for the visibility-based MROP approach. 
Secondly, in the antenna-based MROP approach, the covariance matrix is never computed as the ROPs are obtained by random beamforming of the antenna measurement vector directly. 
A naive accumulation of projection data across batches thus scales as $\cl O(PB)$ and again a batchwise approach with on-the-fly compression will automatically reduce this to $\cl O(PM)$. 
Fig.~\ref{fig:acquisition}(b) illustrates the batchwise implementation for the antenna-based MROP approach. 

\noindent\textbf{Batchwise computation at reconstruction.}
As exemplified in Section~\ref{sec:usara} for the uSARA algorithm, image reconstruction algorithms are iterative, and each iteration involves an update of the image estimate to enforce data fidelity on the one hand and an update enforcing the image model on the other hand. Implementing the computation of the imaging models as the consecutive application of their operators, in particular passing by the $VB$ visibilities, implies that a memory requirement $\cl O(\max(VB,N/2))$, where the factor $1/2$ accounts for the fact that, while the $VB$ visibilities are complex-valued, the $N$ pixel intensities are real-valued.
However, deploying \emph{batchwise} compression and summing the compressed data on the fly over batches, enables reducing the memory requirements, now scaling as $\cl O(\max(V,\alpha D,N/2))$. This process is very similar to the batchwise approach at acquisition, and is illustrated for the MROP model in Fig.~\ref{fig:acquisition}(d). 

\noindent\textbf{Precomputation-free forward operators at reconstruction.}
Reducing the cost of the forward operator of any model is possible at the price of precomputing the composed forward operator.
However, storing the precomputed operator in matrix form significantly increases memory requirements. 
We thus assume here that all computations are performed on the fly, generating the forward operators when needed from a minimal amount of information. In this context, the memory requirements for the forward operators of all sensing models can be neglected compared to the size of the intermediate and final data. 
Firstly, there are no memory requirements for the NUFFT operator $\bsmat{GFZ}$.
Indeed, it can be computed on the fly as it is fully determined by the positions of the $Q$ antennas and the Earth's dynamics. 
Secondly, the compression operators associated with the random Gaussian compression model and MROP, CROP, and IROP can be retrieved with negligible cost by storing random seeds. This avoids having to store the random vectors in memory.

\subsection{Illustrative Numerical Setup} \label{sec:tab1}

Table~\ref{tab:values} reports illustrative numerical values borrowed from the numerical experiments that follow in Sections~\ref{sec:setting}-\ref{sec:results}. 
These will be useful reference points for the complexity analysis. 
In this scenario, we thus consider the reconstruction of a monochromatic intensity image with $N=256\times 256 \simeq 7~10^4$ pixels, whose pixel intensities are real-valued. We consider an array with $Q=64$ dishes such as MeerKAT, which yields $V=Q(Q-1)/2=2016$ baselines. 
We assume $I=8~10^9$ data samples per batch to compute the sample covariance matrix, and $B=2.7~10^3$ batches of duration $IT=8$s, which corresponds to a total acquisition of 12 hours.
The number of visibilities provided by BDA is given by $VB_{\rm bda}\simeq 0.4VB$ for an average number of batches per baseline given by $B_{\rm bda}=1.1~10^3 \simeq 0.4B$. With regards to the MROP model, the simulations deployed in Sections~\ref{sec:setting}-\ref{sec:results} will establish that the number of MROPs needed to preserve the imaging quality of the classical and BDA models is around the image size rather than the number of visibilities, \ie $PM \simeq N \ll VB_{\rm bda}<VB$. 
More specifically, the MROP setup considered takes $(P,M)=(80,800)$, leading to $PM \simeq 6~10^4$. The experiments will establish that the simpler CROP model does in practice require $P_{\rm crop}=48 $, leading to $P_{\rm crop}B\simeq 10^5$, to preserve the imaging quality. 
The number of nearest neighbours used for the NUFFT is  $7$ along each dimension of the image, giving $J=49$.

\begin{table}
    \centering
    \caption{Illustrative numerical setup. Values are borrowed from the numerical experiments that follow in Sections~\ref{sec:setting}-\ref{sec:results}, with the exception of the random Gaussian projection and IROP cases, where values are loose estimates.}
    \label{tab:values}
    \begin{tabular}{|l|r|l|} 
        \hline
        \textbf{Symbol} & \textbf{Value} & \textbf{Number of} \\
        \hline
        $N$ & $256^2$ & Image pixels \\
        \hline
        $Q$ & $64$ & Antennas \\
        \hline
        $V$ & $2016$ & Baselines \\ 
        \hline
        $I$ & $8~10^9$ & Samples per batch \\
        \hline
        $T$ & $10^{-9}$ & Sampling period \\
        \hline
        $B$ & $2700$ & Batches \\ 
        \hline
        $B_{\rm bda}$ & $1100$ & Batches after BDA \\
        \hline
        $P$ & $80$ & Projections in MROP \\
        \hline
        $M$ & $800$ & Modulations \\
        \hline
        $P_{\rm crop}$ & $48$ & Projections in CROP \\
        \hline
        $\cl P$ & \cellcolor{Mylightgrey} $3~10^4$ & Gaussian projections \\
        \hline
        $P_{\rm irop}$ & \cellcolor{Mylightgrey} $6~10^4$ & Projections in IROP \\
        \hline
        $J$ & $49$ & Neighbours for interpolation with $\bsmat{G}$ \\
        \hline
    \end{tabular}
\end{table}

\begin{table}
    \centering
    \caption[Acquisition cost]{Computational cost and memory requirement of the acquisition models in $\cl O$ notation. 
    Injecting the example values of Table~\ref{tab:values} illustrates the order of magnitude with the blue and red colors indicating low and high values, respectively. Cells with numerical values are highlighted in grey background for the random Gaussian projection and IROP cases, as these values are loose estimates of the number of required projections, rather than experimentally validated values as for all other models. All numerical values are rounded up to the first significant digit.
    }
    \label{tab:acquisition}
    \resizebox{\linewidth}{!}{
    \begin{tabular}    {|r|l|l|r|l|r|}
        \hline
        \textbf{Model} & \textbf{Name} & \multicolumn{2}{c|}{\textbf{Acquisition cost}} & \multicolumn{2}{c|}{\textbf{Memory requirement}} \\
        & & \multicolumn{2}{c|}{} & \multicolumn{1}{l}{$\max(V,\alpha D)$} & \\
        \hline
        $\acqop$ & Classical & $IVB$ & \textcolor{Myblue}{$4~10^{16}$} & $VB$ & \textcolor{Myred}{$5~10^6$} \\
        $\acqop^* \acqop$ & Dirty & $IVB+JVB$ & \textcolor{Myblue}{$4~10^{16}$} & $\max(V,N/2)$ & \textcolor{Myblue}{$3~10^4$} \\ 
        $\bsmat{S}\acqop$ & BDA & $IVB$ & \textcolor{Myblue}{$4~10^{16}$} & $VB_{\rm bda}$ & \textcolor{Myred}{$2~10^6$} \\ 
        $\bsmat A \acqop$ & Gaussian & $IVB + \cl PVB$ & \cellcolor{Mylightgrey} \textcolor{Myblue}{$4~10^{16}$} & $\max(V,\cl P)$ & \cellcolor{Mylightgrey} \textcolor{Myblue}{$3~10^4$} \\ 
        $\acqopn$ & CROP (Ant.) & $2IP_{\rm crop}QB$ & \textcolor{Myblue}{$9~10^{16}$} & 
        $\max(V,P_{\rm crop}B)$ & \textcolor{Myblue}{$10^5$} \\
        $\bsmat D \acqop$ & CROP (Vis.) & $IVB + 2P_{\rm crop}VB$ & \textcolor{Myblue}{$4~10^{16}$} & $\max(V,P_{\rm crop}B)$ & \textcolor{Myblue}{$10^5$} \\ 
        $(\bs 1_B \otimes \Id) \acqopn$ & IROP (Ant.) & $2IP_{\rm irop}QB$ & \cellcolor{Mylightgrey} \textcolor{Myred}{$2~10^{19}$} & 
        $\max(V,P_{\rm irop})$ & \cellcolor{Mylightgrey} \textcolor{Myblue}{$6~10^4$} \\
        $\bsmat{R}\acqop$ & IROP (Vis.) & $IVB + 2P_{\rm irop}VB$ & \cellcolor{Mylightgrey} \textcolor{Myblue}{$4~10^{16}$} & $\max(V,P_{\rm irop})$ & \cellcolor{Mylightgrey} \textcolor{Myblue}{$6~10^4$} \\ 
        $\bsmat M\acqopn$ & MROP (Ant.) & $2IPQB + PMB$ & \textcolor{Myblue}{$2~10^{17}$} & 
        $\max(V,PM)$ & \textcolor{Myblue}{$6~10^4$} \\
        $\bsmat{MD} \acqop$ & MROP (Vis.) & $IVB + 2PVB + PMB$ & \textcolor{Myblue}{$4~10^{16}$} & $\max(V,PM)$ & \textcolor{Myblue}{$6~10^4$} \\
        \hline
    \end{tabular}
    }
\end{table}

\begin{table}
    \centering
    \caption[Complexities of the image reconstruction]{Computational cost and memory requirements of the imaging models in $\cl O$ notation. 
    Injecting the example values of Table~\ref{tab:values} illustrates the order of magnitude with the blue and red colors indicating low and high values, respectively. Cells with numerical values are highlighted in grey background for the random Gaussian projection and IROP cases, as these values are loose estimates of the number of required projections, rather than experimentally validated values as for all other models. All numerical values are rounded up to the first significant digit.
    }
    \label{tab:imaging}
    \resizebox{\linewidth}{!}{
    \begin{tabular}{|r|l|l|r|l|r|}
       \hline
        \textbf{Model} & \textbf{Name} & \multicolumn{2}{c|}{\textbf{Computational cost}} & \multicolumn{2}{c|}{\textbf{Memory requirement}} \\
        & & \multicolumn{1}{l}{$2N\log N$} & \textcolor{Myblue}{$2~10^6$} & \multicolumn{1}{l}{$\max(V,\alpha D,N/2)$} & \\
        & & \multicolumn{1}{l}{$+$} & $+$ & \multicolumn{2}{c|}{} \\
        \hline
        $\imop$ & Classical & $JVB$ & \textcolor{Myblue}{$3~10^8$} &   $\max(VB,N/2)$ & \textcolor{Myred}{$5~10^6$} \\  
        $\imop^* \imop$ & Dirty & $2JVB$ & \textcolor{Myblue}{$6~10^8$} &  $\max(V,N/2)$ & \textcolor{Myblue}{$3~10^4$} \\
        $\imop_{\rm bda}$ & BDA &  $JVB_{\rm bda}$ & \textcolor{Myblue}{$10^8$} & 
       $\max(VB_{\rm bda},N/2)$ & \textcolor{Myred}{$2~10^6$} \\ 
       $\bsmat A \imop$ & Gaussian & $JVB+\cl PVB$ & \cellcolor{Mylightgrey} \textcolor{Myred}{$2~10^{12}$} 
        & $\max(V,N/2)$ & \cellcolor{Mylightgrey} \textcolor{Myblue}{$3~10^4$} \\
        $\bsmat{D}\imop$ & CROP & $JVB+2P_{\rm crop}VB$ & \textcolor{Myblue}{$8~10^8$} & $\max(V,P_{\rm crop}B,N/2)$ & \textcolor{Myblue}{$10^5$} \\
        $\bsmat{R}\imop$ & IROP & $JVB+2P_{\rm irop}VB$ & \cellcolor{Mylightgrey} \textcolor{Myred}{$6~10^{12}$} & $\max(V,P_{\rm irop},N/2)$ & \cellcolor{Mylightgrey} \textcolor{Myblue}{$6~10^4$} \\
        $\bsmat{MD}\imop$ & MROP &  $JVB+2PVB+PMB$ & \textcolor{Myblue}{$10^9$} & $\max(V,PM,N/2)$ & \textcolor{Myblue}{$6~10^4$} \\ 
        \hline
    \end{tabular}
    }
\end{table}

As already mentioned, Sections~\ref{sec:comp_classic}-\ref{sec:comp_mrop} will establish that the random Gaussian projection and IROP models are non-viable as they would entail too high a computational cost, on which basis they are discarded from the numerical experiments presented in Sections~\ref{sec:setting} and \ref{sec:results}. 
In this context, while no explicit numerical value is available for their number of projections, loose estimates are discussed in the complexity analysis, with $\cl P \lesssim \cl O(N/2)$ for the Gaussian projection model and $P_{\rm irop} \sim \cl O(PM)$ for IROP. These relations are used to set illustrative values as $\cl P = N/2= 3~10^4$ and $P_{\rm irop}=PM= 6~10^4$.

For readability of the complexity analysis that follows, we adopt a notation consisting in appending a suffix to a complexity to highlight the value it takes in our illustrative numerical setup, \eg $\cl O(JVB)_{\rightarrow 3~10^8}$. We also highlight that all numerical values in $\cl O$ notation are rounded up to the first significant digit.

\subsection{Classical Model} \label{sec:comp_classic}

\textbf{Acquisition cost.}
To acquire the visibilities $\bs v = \acqop(\cl X)+\bs\xi$, the (upper triangular part of the) outer product of each antenna signal vector $\bs x_b[i]$, with $i\in \upto{I}$, is taken, costing $\cl O(V)_{\rightarrow 2~10^3}$ operations per time instant, hence giving a total cost of $\cl O(IVB)_{ \rightarrow 4~10^{16}}$.

\noindent \textbf{Data size and memory requirement at acquisition.} Both the final data size and intermediate memory requirements are $\cl O(VB)_{\rightarrow 5~10^6}$, driven by the total number of visibilities, which is the very motivation for dimensionality reduction.

\noindent \textbf{Computational cost at reconstruction.} The classical imaging model \eqref{eq:classic} consists of the consecutive application of $\bsmat{Z}$, $\bsmat{F}$ and $\bsmat{G}$, giving a computational cost $\cl O(2N \log N + JVB)_{\rightarrow 3~10^8}$, where the $2N \log N$ term comes from a real FFT applied on a $4N$-sized vector, and the $JVB$ term is for the application of $\bsmat{G}$.
In the illustrative numerical setup, the cost is dominated by the interpolation with $\bsmat{G}$.

\noindent \textbf{Memory requirement at reconstruction.}
The memory requirement at reconstruction is $\cl O \big( \max(VB,N/2) \big)_{\rightarrow 5~10^6}$. This value is dominated by $VB$ in our numerical setup .

\subsection{Dirty Image Model} \label{sec:comp_dirty}

\noindent\textbf{Acquisition cost.} 
The dirty image can be acquired in a batchwise manner as $\ddirty = \real{\sum\btoB \imop_b^* \acqop_b(\cl X)}$.
The total acquisition cost is given by the computation of the visibilities at each time instant plus their back-projection in each batch, \ie $\cl O(IVB+JVB) \simeq \cl O(IVB)_{\rightarrow 4~10^{16}}$.
An antenna-based method, named EPIC, exists to acquire the dirty image directly from the signal set $\cl X$ as $\ddirty = \real{\acqop^* \acqop(\cl X)}$ \citep{thyagarajan17}. 
This acquisition method has the advantage to depend linearly on $Q$. 
Indeed, its computation costs $\cl O(JQ + N_b' \log N_b')_{\rightarrow 8~10^4}$ per time instant and $\cl O(I(JQ + N_b' \log N_b')B)_{\rightarrow 2~10^{18}}$ in total, where $N_b'$ is the average number of gridded visibilities per batch (\ie the number of $uv$ cells activated by the visibilities of one batch, corresponding to the number of nonzero columns in the interpolation operator $\bsmat G_b$, as illustrated in Fig.~\ref{fig:operators}), which takes a value $6~10^{3}$ in our illustrative setup. This alternative approach was proposed for dense arrays, \ie for large $Q$ only, and indeed appears suboptimal in our illustrative numerical setup.

\noindent \textbf{Data size and memory requirement at acquisition.} The memory requirement for the final data of the dirty image model is $\cl O(N/2)$. This approach represents an efficient data dimensionality reduction as soon as $N/2 \ll VB$. Intermediate memory requirements scale as $\cl O(\max(V,N/2))$, which could also be dominated by $V$ in the case of dense arrays.
Alternatively, working with the $ N' < 4N$ gridded visibilities obtained as $\bsmat{G}^*\bs v$ may provide further savings if $ N' < N/2$. 
In our numerical setup, $N'=6~10^4$ and the memory requirement is lower for the ``dirty image'' approach, as opposed to the ``gridded visibilities'' approach.
For the sake of simplicity, Table \ref{tab:acquisition} provides the memory requirements for the former rather than the latter. 

\noindent \textbf{Computational cost at reconstruction.}
The cost of computing the imaging model $\bar{\bs\sigma}_\rmd = \real{\imop^* \imop(\bs\sigma)}$ is twice the cost of the classical model, hence having a complexity order $2\cl O(2N\log N + JVB)_{\rightarrow 6~10^8}$.

\noindent \textbf{Memory requirement at reconstruction.}
The memory requirement simply reads $\cl O\big(\max(V,N/2)\big)_{\rightarrow 3~10^4}$. In our illustrative setup, the image size is dominant.

\subsection{BDA} \label{sec:comp_bda}

\textbf{Acquisition cost.} The acquisition cost is the same as for the classical model, requiring $\cl O(IVB)_{\rightarrow 4~10^{16}}$ operations, as the additional baseline-dependent averaging entails negligible overhead smaller than $\cl O(VB)$.

\noindent \textbf{Data size and memory requirement at acquisition.}
Averaging the visibilities (in a batchwise manner)  reduces intermediate memory requirements and data size to $\cl O(VB_{\rm bda})_{\rightarrow 2~10^6}$ as opposed to $\cl O(VB)_{\rightarrow 5~10^6}$ for the classical model.

\noindent \textbf{Computational cost at reconstruction.}
The forward model has the same structure as the classical model, but requires the computation of $B_{\rm bda}<B$ visibilities, thus reducing computational cost to $\cl O(2N \log N + JVB_{\rm bda})_{\rightarrow 10^8}$ operations.

\noindent \textbf{Memory requirement at reconstruction.}
During image reconstruction, the memory requirement is given by $\cl O\big(\max(VB_{\rm bda},N/2)\big)_{\rightarrow 2~10^6}$. 
Like for the classical model, this value is dominated in our setup by the number of visibilities.

\subsection{Random Gaussian compression} \label{sec:comp_gaussian}

Such approach has been exhaustively demonstrated by the theory of compressive sensing to deliver optimal data dimensionality reduction \cite{foucart17}, with the number of projections needed scaling as the sparsity $K$ of the image model, and in all instances upper-bounded by the image dimension $\cl P \lesssim \cl O(N/2)$. 

\noindent \textbf{Acquisition cost.}
The sample covariance matrices are acquired for $\cl O(V)_{\rightarrow 2~10^3}$ operations per time instant, and a number $\cl P$ of projections of these sample covariance matrices are computed for each batch. 
The total acquisition cost thus scales as $\cl O(IVB + \cl PVB) \simeq \cl O(IVB)_{\rightarrow 4~10^{16}}$.

\noindent\textbf{Data size and memory requirement at acquisition.}
While the memory requirement $\cl O(\cl P)_{\rightarrow 3~10^4} $ for the data is optimal, the intermediate memory requirement $\cl O(\max(V,\cl P))_{\rightarrow 3~10^4}$ may well be dominated by the number of baselines for dense arrays, even if it is not the case in our numerical setup.

\noindent \textbf{Computational cost at reconstruction.}
The imaging model \eqref{eq:gaussian_im} has a total cost $\cl O(2N \log N + JVB + \cl PVB)_{\rightarrow 2~10^{12}}$. 
This will be dominated by the random projection term $\cl O(\cl PVB)$ as soon as $\cl P > J$. Unless very simple images are considered, with very small sparsity,  $\cl P$ will not be far from its upper bound $N/2$, leading to highly unaffordable computational costs. This explains why such approach rapidly becomes impractical and alternative dimensionality reduction techniques are required \citep{kartik17}.

\noindent \textbf{Memory requirement at reconstruction.}
Assuming batchwise computation of the projections, the largest intermediate data size is given by $\cl O\big(\max(V, N/2)\big)_{\rightarrow 3~10^4}$. 
In the illustrative setup, the $N/2$ term dominates.

\subsection{MROP} \label{sec:comp_mrop}

This section compares the computational costs and memory requirements of both the antenna- and visibility-based ROP models, analysing in detail (i) the general MROP proceeding to random projections followed by random modulations, (ii) the CROP variant where the random projections are not modulated, and (iii) the IROP variant where the random projections are summed up. 

\noindent\textbf{Acquisition cost.}
As discussed in Section~\ref{sec:mrop1} and illustrated in Fig.~\ref{fig:acquisition}(b), the antenna-based acquisition of MROPs does not compute the covariance matrices. 
Instead, it requires $\cl O(2PQ+P)$ operations for the projections with $\bc A$ and $\bc B$ and for the product in \eqref{eq:proj_prod} at each time instant.
The computational cost of the random modulations is $\cl O(PMB)$. 
The resulting total acquisition cost is $ \cl O(2IPQB+PMB) \simeq \cl O(2IPQB)_{\rightarrow 2~10^{17}}$, which is smaller than the classical sensing scheme if $2PQ \leq V$ and larger otherwise. 
The CROP model will incur the same complexity but with $P\rightarrow P_{\rm crop}$ and $M\rightarrow 0$. Acknowledging that $P_{\rm crop}\leq P$ in general as there is no subsequent compression via random modulation, offers a potential advantage to CROP over MROP. In our illustrative setup, $P_{\rm crop}= 0.6 P$, confirming the advantage is realised in practice. 
In the IROP case, $P\rightarrow P_{\rm irop}$ and $M\rightarrow 1$ as a simple summation is applied. The results in \citet{leblanc2024b} suggest that the product $PM$ is to some extent the driving factor for the quality of image reconstruction in MROP, rather than $P$ and $M$ separately. In the asymptotic case where $M\rightarrow 1$, one thus have  $ P_{\rm irop} \sim \cl O(PM)$, leading to a highly unaffordable acquisition cost $\cl O(2IP_{\rm irop}VB)_{\rightarrow 2~10^{19}}$. 

The visibility-based acquisition of MROP is a post-correlation scheme, meaning the covariance matrix must be computed batchwise at a cost $\cl O(V)_{\rightarrow 2~10^3}$. 
Then, the ROPs must be explicitly computed for each batch in $\cl O(2PV)$ operations.
The computational cost of the random modulations is $\cl O(PMB)$.
The resulting total acquisition cost is thus $\cl O(IVB+2PVB+PMB) \simeq \cl O(IVB)_{\rightarrow 4~10^{16}}$, which is the same as the classical sensing scheme. 
As the $O(IVB)$ term dominates largely, the CROP and IROP models will be driven by the same computational cost.

In our illustrative numerical setup, the antenna-based approach for MROPs is 5 times more costly than the visibility-based approach and classical model. 
However, for dense arrays, if $PM$ is indeed the driving factor, the antenna-based setting offers an opportunity to bring the MROP 
acquisition cost below the requirements of the classical scheme by increasing $M$ so that $2PQ \leq V$. 
Validating this statement requires further investigations beyond the scope of the current work.

\noindent \textbf{Data size and memory requirement at acquisition.}
The memory requirement for storing the final MROP data is $\cl O(PM)_{\rightarrow 6~10^4}$. 
In our numerical setup $PM \simeq N$, so that the MROP model provides a highly efficient data dimensionality reduction with respect to the classical model when the number of visibilities $VB$ is much larger than the image size, of the same order of (but slightly suboptimal compared to) the reduction to $O(N/2)$ delivered by the dirty image model. 
The intermediate memory requirements are  $\cl O\big(\max(V, PM)\big)_{\rightarrow 6~10^4}$. 
In the illustrative setup, the $PM$ term dominates. 

With regards to the CROP model, the memory requirement for data storage is $\cl O(P_{\rm crop}B)_{\rightarrow 1~10^5}$. 
Whether it compares favourably or not to MROP's $O(PM)$ depends on how much below $P$ the value $P_{\rm crop}$ can be chosen. In our illustrative setup CROP requires twice more memory than MROP: $P_{\rm crop}B=2PM$. 
With regards to the IROP model, the fact that the required $P_{\rm irop}$ may still be smaller than MROP's $PM$ is a potential advantage in terms of memory requirement, scaling as $\cl O(P_{\rm irop})_{\rightarrow 6~10^4}$. 
In the illustrative setup, IROP is on par with MROP.

\noindent \textbf{Computational cost at reconstruction.}
The operators involved in the MROP imaging model (see Section~\ref{sec:mrop1}) are illustrated in and Fig.~\ref{fig:acquisition} and Fig.~\ref{fig:operators}. 
The imaging cost is given by $\cl O(2N\log N + JVB + 2PVB + PMB)_{\rightarrow 10^9}$. 
In general, discarding the FFT term on the assumption of a large enough number of visibilities $VB$, if $2P>J$ and $M>V$, the $\cl O(PMB)$ cost of the modulations will dominate. 
However, as these modulations imply flipping signs and complex summations in contrast to complex multiplications, their unit cost is generally smaller than the unit cost of the other terms, so that it will take even larger $P$ and $M$ values for this term to dominate.
If $2P>J$ and $M<V$, the $O(2PVB)$ cost of ROPs dominates, which is the case in our numerical setup. 
Only when $2P<J$ and $M<V$ would the cost reduce down to that of the classical model. 

For the CROP model, $P\rightarrow P_{\rm crop}$ and $M\rightarrow 0$, leading to a computational
cost $\cl O(2N \log N + JVB + 2P_{\rm crop}VB)_{\rightarrow 8~10^8}$ and placing CROP in an even better position than MROP to reduce the computational cost down to that of the classical model. In the illustrative setup, CROP indeed entails a slightly smaller complexity as $P_{\rm crop}=0.6 P$. 
For the IROP model, $P\rightarrow P_{\rm irop}$ and $M\rightarrow 1$, leading to the highly unaffordable computational
cost $\cl O(2N \log N + JVB + 2P_{\rm irop}VB)_{\rightarrow 6~10^{12}}$ if $P_{\rm irop} \sim \cl O(PM)$, just as for random Gaussian compression.

\noindent \textbf{Memory requirement at reconstruction.}
Assuming batchwise computation, MROP's memory requirement is set by $\max(V,PM,N/2)_{\rightarrow 6~10^4}$.
In the numerical setup, the $PM$ term dominates by a factor 2 over the $N/2$ term as $PM\simeq N$. 
The memory requirement for CROP scales as $\max(V,P_{\rm crop}B,N/2)$, which compares favourably to MROP only if $ P_{\rm crop}B\leq PM$. This condition is not met in our numerical setup as $P_{\rm crop}B=2 PM$, and CROP is suboptimal. 
The memory requirement for IROP scales as $\max(V,P_{\rm irop},N/2)_{\rightarrow 6~10^4}$, similar to MROP if $ P_{\rm irop} \sim \cl O(PM)$. 
In the illustrative setup, IROP is thus on par with MROP.

\subsection{First Conclusions on Viable Models} \label{sec:firstconclusion}

With regards to memory requirement, MROP and CROP offer a great opportunity for a significant reduction in data size and associated memory requirements with respect to the classical and BDA models. With regards to computational cost at acquisition and reconstruction, MROP and CROP can enable imaging at a complexity similar to the classical model.
IROP will be as unaffordable as the pure Gaussian random projection approach. 
In this context, neither IROP nor the Gaussian random projection models are considered viable, and they are discarded in the numerical experiments presented in Sections~\ref{sec:setting}-\ref{sec:results}. 

The elephant in the room in this context is the dirty image model, or the related visibility-gridding model. 
Section \ref{sec:dirty_remark} discusses why it does not represent the ultimate solution to dimensionality reduction and is therefore also discarded from our numerical experiments, leaving only the classical model and BDA as competitors to MROP and CROP.

\subsection{Blessing and Curse of Dirty Image Model} \label{sec:dirty_remark}

The classical model iterations of gradient-based image reconstruction algorithms, such as uSARA used in this paper and presented in Section~\ref{sec:usara}, only rely on $\imop^* \imop$ and the dirty image $\ddirty$ for the image reconstruction process. 
In such an algorithm-specific context, the acquisition cost, memory requirement, and imaging cost of the classical model naturally boil down to those of the dirty image model. 
This represents a distinct advantage of the dirty image model, which brings the memory requirements of the classical model down from data scale $\cl O(VB)$ to image scale $\cl O(N/2)$. 

However, in RI, the problematic of the calibration of antenna-based direction-independent effects (DIEs) and direction-dependent effects (DDEs) is integral to the image formation process. 
The general self-calibration approach consists in alternating image estimation and DIE/DDE estimation loops. 
This self-calibration process can be formulated as a joint inverse problem for the image and calibration variables, and methods have been developed for a provably convergent process \citep{repetti2017non,dabbech2021cygnus}. 
In this context, it is paramount to ensure that any data dimensionality reduction approach remain ``calibration-friendly'', which can be ascertained if the dimensionality reduction operator is linear and independent of the image and calibration variables themselves. 
Under such constraint, DIEs/ DDEs can simply be introduced in the measurement model as additional variables before the dimensionality reduction operator.

All compression schemes considered here do satisfy these constraints with the notable exception of the dirty image approach, which introduces further coupling between the calibration variables by using $\imop^*$ to reduce the original data dimension down to image dimension. 
As a result, in the wider context of joint self-calibration and imaging, the dirty image approach raises significant challenges, calling for alternative approaches. The dirty image model will thus not be considered in the numerical experiments presented in Sections~\ref{sec:setting} and \ref{sec:results}.
\section{Simulations} \label{sec:setting}

The remainder of this paper aims to demonstrate the interest of the MROP scheme described in Section~\ref{sec:compression} as an efficient compression technique, in terms of computational efficiency and image reconstruction performance. 
With that aim, this section presents a realistic numerical setting for an extensive comparative study between the classical, subsampled, BDA, CROP, and MROP models using the image reconstruction algorithm uSARA.  

\subsection{Ground-truth Images} \label{sec:gt}

\begin{figure}
    \centering
    \includegraphics[width=1\linewidth]{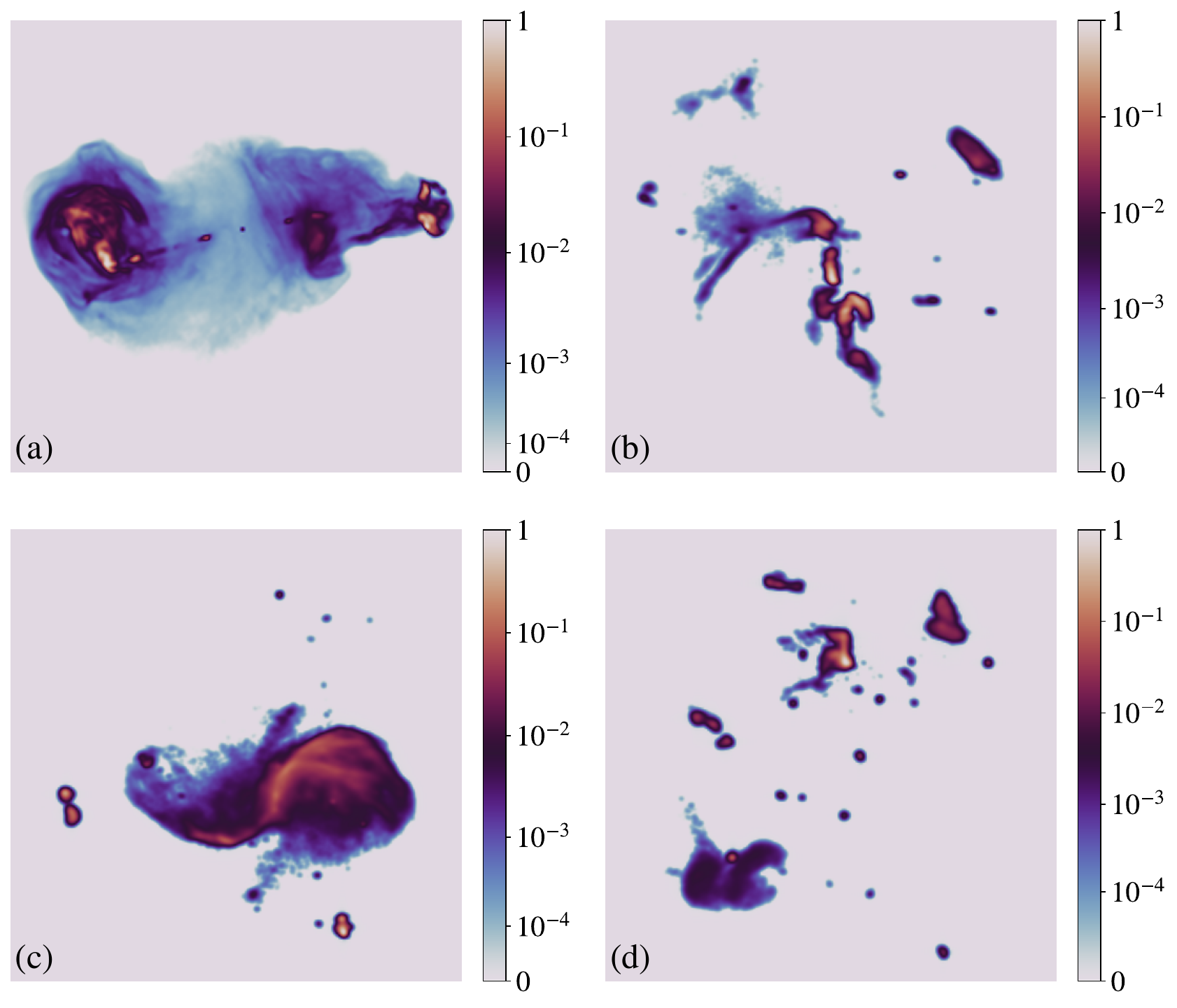}
    \caption{Ground-truth images made of $256\times 256$ pixels and shown in logarithmic scale. (a) 3c353, (b) Abell 2034, (c) Messier 106, (d) PSZ2 G165.68+44.01. }
    \label{fig:groundtruth}
\end{figure}

In the following experiments, we consider four ground-truth radio images shown in Fig.~\ref{fig:groundtruth}, namely the radio galaxies 3c353 and Messier 106, and the galaxy clusters Abell 2034 and PSZ2 G165.68+44.01, all resized to $N=256^2$ pixels.
The dynamic range of each ground-truth image, denoted by $d_0$, is defined as the ratio between the maximum and faintest pixel intensities. Since all images are normalised to have a peak value of $1$, their dynamic range is the reciprocal of the faintest intensity, denoted by $\bar{\tau}_0 = d_0^{-1}$, which is (somewhat arbitrarily) computed as the fifth percentile of the nonzero pixel intensity distribution.

For robust validation, we target the reconstruction of these ground-truth images with different dynamic range values $d = \bar{\tau}^{-1} > d_0$.
In that aim, we use the exponentiation operation introduced in \citet{terris22}, defined as
\begin{equation} \label{eq:exponentiation}
    \bs u_{\rm e} := (\zeta^{\bs u}-1)/\zeta,
\end{equation}
where $\bs u$ is the original image, $\bs u_{\rm e}$ indicates its exponentiated version, and $\zeta$ is the exponentiation factor given by the numerical solution of  
\begin{equation} \label{eq:solve_expo_factor}
    \zeta = (1+\zeta \bar{\tau}^{-1})^{\bar{\tau}_0}.
\end{equation}
In our experiments, 25 values of dynamic range sampled as $\log_{10} (d^{-1}) \sim \cl U[-5, -3]$ are assigned to each of the 4 ground-truth images, resulting in 100 unique ground-truth images covering a wide spectrum of dynamic ranges, \ie $d\in [10^3, 10^5]$.

\subsection{Sensing Models} \label{sec:sensing_models}

As we target a comparison between different sensing models rather than absolute evaluations, we will simplify our numerical analysis by considering the reconstruction of radio images of galaxies sensed through the forward imaging models only. In other words, we do not consider the real acquisition process nor the gap between this acquisition and the imaging process, which is mainly modelled by an additional (statistical) noise term and generally small compared to other noise sources owing to a high number of samples per batch $I$ \citep{thompson80vla}.

The ground-truth image will be sensed through  
the classical model \eqref{eq:classic}, the subsampled model, the BDA model \eqref{eq:baseline_dep_av_im}, and the CROP \eqref{eq:sep_rop2} and MROP models \eqref{eq:im_mrop}.
The subsampled model is used as a trivial benchmark, placing a lower bound on the image reconstruction quality with any more sophisticated compression model. 

In all five cases, \iid complex Gaussian noise $\bs\xi \sim \cl{CN}(0,\tau^2 \Id)$ is added to the visibilities as in \eqref{eq:classic}, so that the additional compression operators $\bsmat P$ for subsampled, $\bsmat S$ for BDA, $\bsmat{D}$ for CROP, and $\bsmat{MD}$ for MROP, are applied to that noise. 
Following \citet{terris22}, the standard deviation of $\bs\xi$ is modelled as $\tau = \bar{\tau} \sqrt{2L}$ where $L:=\norm[\big]{\real{\imop^{*} \imop}}{\rm S}$ is the \emph{Lipschitz constant}, or the spectral norm of the classical operator $\imop$. This ensures that the ``target dynamic range of reconstruction'' (computed as $\tau^{-1} \sqrt{2L}$, the reciprocal of the standard deviation of the back-projected noise $\real{\imop^{*}\bs \xi}^{-1}$) is equal to the dynamic range of the ground truth $\bar{\tau}^{-1}$. 
The five resulting forward imaging models are
\begin{align} 
    \bs v_{~} & = \bsmat{W}\imop(\bs\sigma) + \bsmat{W}\bs\xi, \label{eq:classic2} \\
    \bs v_{\rm sub} & = \bsmat{W}_{\rm sub}\imop_{\rm sub}(\bs\sigma) + \bsmat{W}_{\rm sub}\bsmat{P}\bs\xi, \label{eq:sub2} \\
    \bs v_{\rm bda} & = \bsmat{W}_{\rm bda}\imop_{\rm bda}(\bs\sigma) + \bsmat{W}_{\rm bda}\bsmat{S}\bs{\xi}, \label{eq:bda2} \\
    \bs y_{~} & = \bsmat{DW}\imop(\bs\sigma) + \bsmat{DW}\bs\xi, \label{eq:crop2} \\
    \bs z_{~} & = \bsmat{MDW}\imop(\bs\sigma) + \bsmat{MDW}\bs\xi, \label{eq:mrop2}
\end{align}
where $\bsmat P \in \{ 0,1\}^{VB_{\rm sub}\times VB}$ is a selection matrix subsampling the visibilities across the batches, and $\imop_{\rm sub} := \bsmat{P}\imop$ is the \emph{subsampled} forward operator. 
Our proposed simulation setup focuses on uniform visibility weighting for $\bsmat{W} \in \Rbb^{VB\times VB}$, $\bsmat{W}_{\rm sub} \in \Rbb^{VB_{\rm sub} \times VB_{\rm sub}}$, and $\bsmat{W}_{\rm bda} \in \Rbb^{VB_{\rm bda} \times VB_{\rm bda}}$,  where $B_{\rm sub}$ and $B_{\rm bda}$ are the numbers of subsampled and BDA-ed batches. The BDA model follows a deterministic averaging procedure described in \cite{wijnholds18}, and the averaged $uv$-point of the corresponding averaged visibilities are chosen to form $\bsmat G_{\rm bda}$, as described in Section~\ref{sec:post_corr}. 

\subsection{$uv$-coverages and Data Sizes} 
\label{sec:uv_and_size}

\begin{figure}
\centering  \includegraphics[width=1\linewidth]{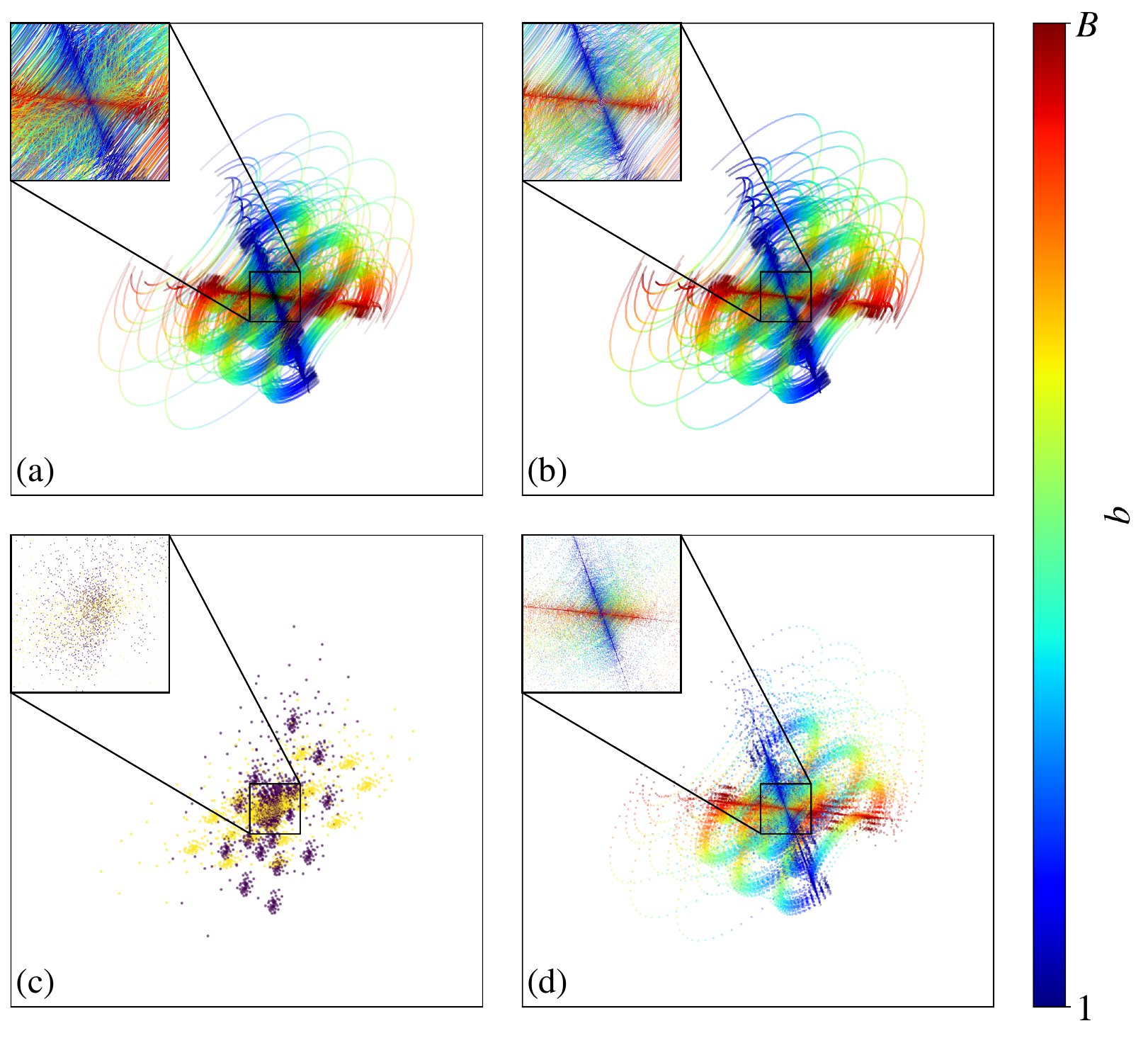}
\caption{
One of the $uv$-coverages of simulated MeerKAT observations used within our experiments, with a super-resolution factor of 1.5. 
(a) classical, (b) averaged $uv$ points with BDA, (c) subsampled $uv$ points with $B_{\rm bda} = 2$, (d) subsampled $uv$ points with $B_{\rm bda}=32$. Each panel is accompanied by a zoom-in of the $uv$-coverage, showing baselines shorter than 1 km in its top left corner.
For BDA, the central batch's colour is selected from the available averaged colours.
Furthermore, concentric circles appear around the centre, depicting different averaging regimes.
Each subplot lives in $[0,1) \times [0,1)$. 
The $uv$-coverage $\cl V_b$ associated with batch $b$ has a unique colour, and the colours are faded from blue for $b=1$ to red for $b=B$.} 
\label{fig:uv}
\end{figure}

Two RI arrays are considered, namely VLA and MeerKAT, with $Q=27$ antennas and $V=351$ baselines for VLA, and $Q=64$ antennas and $V=2016$ baselines for MeerKAT. 
The respective total acquisition times are set as $ITB \in \{21\,640, 43\,232\}$s, and the batch durations as $IT \in \{8, 16\}$s, resulting in batch sizes $B \in \{2\,705, 2\,702\}$. We note that the batch durations were chosen such that its value is small enough for the stationarity assumption below \eqref{eq:xb} to hold, while also large enough so that, with BDA, two visibilities associated to the longest baseline cannot be averaged without affecting reconstruction quality.

For each telescope, $uv$-coverages $\cl V := \bigcup\btoB \cl V_b$ are generated by varying the pointing direction $\bs s_0$, \ie the declination and right ascension of the source. 
The $u$ and $v$ variables can be inferred from \eqref{eq:31_baselines} as the first and second components of the baseline. For the $jk$-th baseline $\bs\nu_{bjk}$, they can be computed as
\begin{equation}
    \begin{bmatrix}
    u \\ v
    \end{bmatrix} = \bs\nu_{bjk} = \frac{\bs p_{bj}^\perp-\bs p_{bk}^\perp}{\lambda}.
\end{equation}
We assume a super-resolution factor of 1.5, defined as the ratio between the maximum spatial frequency and the maximum sampled frequency in either of the two image dimensions.

The data size of the classical model, which represents the \emph{uncompressed} scheme (with no data reduction), is $VB \in \{ 9.5~10^5,5.4~10^6 \}$ for VLA and MeerKAT respectively, leading to \emph{visibility-to-image} ratios $VB/N \in \{14.5, 83.1\}$. 
For the BDA, subsampled, CROP, and MROP models, beyond the data size $D$ (with $D=VB_{\rm bda}$, $D=VB_{\rm sub}$, $D=P_{\rm crop}B$, $D=PM$ respectively), the \emph{data-to-visibility} $D/VB$ and the \emph{data-to-image} $D/N$ compression ratios will be used to assess the level of compression of each model. 
For BDA, the number of averaged visibilities $VB_{\rm bda}$ has no closed-form definition, and is fixed as a function of the number of points and arrangement of the original $uv$-coverage. 
For the chosen numerical setup, the data sizes of BDA are $VB_{\rm bda} \in \{5.4~10^4, 2.3~10^6\}$ for VLA and MeerKAT respectively, and the corresponding ratios are $D/VB \in \{0.570, 0.423\}$ and $D/N \in \{8.26, 35.2\}$.
For the reconstructions using the subsampled model, we will consider the set of batches $B_{\rm sub} \in \{ 3, 12, 26, 72, 183 \}$ for VLA and $B_{\rm sub} \in \{ 1,2,5,13,32\}$ for MeerKAT, leading to the reduced number of visibilities $VB_{\rm sub} \in \{ 1.1~10^3, 4.2~10^3, 9.1~10^3, 2.5~10^4, 6.4~10^4 \}$ for VLA and $VB_{\rm sub} \in \{ 2.0~10^3, 4.0~10^3, 1.0~10^4, 2.6~10^4, 6.5~10^4 \}$ for MeerKAT. This corresponds to the compressions ratios $D/VB \in \{3.7~10^{-4}, 7.4~10^{-4}, 1.9~10^{-3}, 4.8~10^{-3}, 1.2~10^{-2}\}$ and 
$D/N \in \{3.1~10^{-2}, 6.1~10^{-2}, 1.5~10^{-1}, 4.0~10^{-1}, 9.8~10^{-1}\}$ for VLA, and 
$D/VB \in \{1.1~10^{-3}, 4.4~10^{-3}, 9.6~10^{-3}, 2.7~10^{-2}, 6.8~10^{-2}\}$ and 
$D/N \in \{1.6~10^{-2}, 6.4~10^{-2}, 1.4~10^{-1}, 3.9~10^{-1}, 9.8~10^{-1}\}$ for MeerKAT.
Fig.~\ref{fig:uv}(a) depicts an example $uv$-coverage with MeerKAT, while Fig.~\ref{fig:uv}(b) shows the reduced $uv$-coverage after averaging with BDA, and Fig.~\ref{fig:uv}(c) and (d) show the subsampled $uv$-coverages for $B_{\rm sub}=2$ and $B_{\rm sub}=32$, respectively.
For the CROP model, we  set $P_{\rm crop} \in \{1, 2, 3, 9, 24, 48\}$ for both telescopes, leading to the data-to-visibility ratios $D/VB \in \{2.9~10^{-3}, 5.7~10^{-3}, 8.6~10^{-3}, 2.6~10^{-2}, 6.8~10^{-2}, 1.4~10^{-1}\}$ for VLA and $D/VB \in \{5.0~10^{-4}, 9.9~10^{-4}, 1.5~10^{-3}, 4.5~10^{-3}, 1.2~10^{-2}, 2.4~10^{-2}\}$ for MeerKAT, and data-to-image ratios $D/N \in \{4.1~10^{-2}, 8.3~10^{-2}, 1.2~10^{-1}, 3.7~10^{-1}, 9.9~10^{-1}, 2.0\}$ for both.
For the MROP model, as already emphasised in Section~\ref{sec:comp_mrop}, the results in \citet{leblanc2024b} suggest that only the product $PM$ matters for the reconstruction quality. For the sake of simplicity, we consider a single combination of $P$ and $M$ with $P=0.1M$ for each target $PM$ value. 
Technically, we set $(P,M) \in \{ (10,100), (20,200), (30,300), (50,500), (80,800)\}$, corresponding to 
$PM \in \{ 10^3, 4~10^3, 9~10^3, 2.5~10^4, 6.4~10^4 \}$. 
The corresponding data-to-visibility and data-to-image ratios are $D/VB \in \{1.8~10^{-4}, 7.3~10^{-4}, 1.7~10^{-3}, 4.6~10^{-3}, \}$ for VLA,
$D/VB \in \{2~10^{-4}, 7~10^{-4}, 2~10^{-3}, 5~10^{-3}, 1~10^{-2}\}$ for MeerKAT, and
$D/N \in \{1.5~10^{-2}, 6.1~10^{-2}, 1.4~10^{-1}, 3.8~10^{-1}, 9.8~10^{-1}\}$ for both arrays. 

For robust validation we gather image reconstruction statistics over 100 unique inverse problems for VLA and MeerKAT, coupling the 100 ground-truth images (see Section~\ref{sec:sensing_models}) with 100 different $uv$-coverages. 
We will analyse the simulation results in Section~\ref{sec:results} separately for each of the 2 telescopes, also separating the dynamic range into 2 categories defined by $d \in [10^3, 10^4]$ and $d \in [10^4, 10^5]$, leading to about $50$ inverse problems in each category.

\subsection{Reconstruction with uSARA} \label{sec:usara}

Here, we briefly present the gradient-based image reconstruction algorithm used for the experiments of Section~\ref{sec:results}. While simple minimisation algorithms such as \textsc{CLEAN} \citep{CLEAN} and \textsc{BPDN} \citep{pareto} exist, we focus our analysis on a recent state-of-the-art method for high dynamic range RI imaging.
The \emph{unconstrained Sparsity Averaging Reweighted Analysis} (uSARA) \citep{repetti20,terris22} was recently demonstrated to deliver state-of-the-art reconstruction results on the classical model \eqref{eq:classic2}. 
Applied to any of the sensing models \eqref{eq:classic2}-\eqref{eq:mrop2} and calling their forward operator $\bar\imop \in \{ \bsmat{W}\imop, \bsmat{W}_{\rm sub}\imop_{\rm sub}, \bsmat{W}_{\rm bda}\imop_{\rm bda}, \bsmat{DW}\imop, \bsmat{MDW}\imop \}$ and their associated data $\bs b \in \{ \bs v, \bs v_{\rm sub}, \bs v_{\rm bda}, \bs y, \bs z \}$, the minimisation problem associated with uSARA writes 
\begin{equation} \label{eq:usara}
    \wt{\bs\sigma} = \argmin_{\bs u \in \Rbb^N}~\tinv 2 \norm{\bar\imop \bs u-\bs b}{2}^2 + \eta \cl R(\bs u),
\end{equation}
for a measurement data $\bs b \in \Cbb^D$, a \emph{regularisation parameter} $\eta>0$, and prior model 
\begin{equation} \label{eq:prior}
    \cl R(\bs u) := \theta \sum_{j=1}^{9N} \log \big( \theta^{-1} |(\bs\Theta^*\bs u)_j| + 1 \big) + \iota_{\Rbb_+^N}(\bs u),
\end{equation} 
where $\theta>0$ and the $+1$ are used to avoid reaching zero values in the argument of the logarithmic terms. The ``average sparsity'' prior in \eqref{eq:prior} consists of a non-negativity constraint and a log-sum prior (a non-convex approximation of the $\ell_0$-norm) promoting average sparsity in an over-complete wavelet dictionary $\bs\Theta \in \Rbb^{N\times 9N}$. Specifically, the dictionary $\bs\Theta$ concatenates the first eight Daubechies wavelets \citep{daubechies92} and the Dirac basis $\Id_N$. The mean squared error loss $ \norm{\bar\imop \bs u-\bs b}{2}^2/2$ is the considered differentiable data-fidelity term, corresponding to the negative log-likelihood of the data under \iid Gaussian random noise assumption.

The uSARA optimisation problem in \eqref{eq:usara} can be solved using a Forward-Backward algorithmic structure \citep{Beck_book}, involving at each iteration the data-fidelity term via its gradient $\real{\bar\imop^* \bar\imop \bs u - \bar\imop^* \bs b}$ and the prior term via its so-called proximal operator. 
While the algorithmic details and internal parameters of uSARA can be found in \citet{terris22}, the only information needed to run it is the forward operator $\bar\imop$ and the noise level affecting the data, both driving the choice of the parameters $\eta$ and $\theta$. 
A heuristic relation was developed and validated in \citet{terris22} to that effect. 
When the data are affected by \iid Gaussian noise with unit variance, the heuristic takes its simplest form, with $\eta$ and $\theta$ set to $3\theta \simeq 2\eta/L \simeq 1/\sqrt{2L}$. 
When the noise in the data is made non-\iid due to uniform weighting, a correction factor applies, which is discussed in \citet{wilber2023scalable}.
As our simulations focus on uniform weighting, the latter form of the heuristic applies for the classical, subsampled, and BDA sensing models. However, in the case of the CROP and MROP models, we have demonstrated in Section~\ref{sec:mrop_noise} that the randomness of the projections preserves the \iid nature  of the noise even in the presence of uniform weighting, so that the simpler form of the heuristic applies.

In practice, slight adjustments to the heuristic $\theta$, and in turns to the regularisation parameter $\eta$, are necessary to achieve optimal reconstruction quality with a balance between resolution and sensitivity. 
In our experiments, a single heuristic correction is used for all $\sim 50$ inverse problems within a dynamic range category for a given telescope and a given sensing model.

In a reproducibility perspective, we provide explicit stopping criteria for the uSARA algorithm described in \citet[Algo~1-2]{terris22}.
For all tests, the stopping criteria of uSARA must be fixed to $\xi_1 = 10^{-5}$  and $\xi_2 = 10^{-4}$ in Algo~1 and 2, respectively, while limiting the maximum number of iterations in the outer and inner loops of Algo~1 to 3000 and 20, and to 200 in Algo~2, respectively.

\subsection{Implementation} \label{sec:implementation}

Despite the computational considerations discussed in Section~\ref{sec:complexity}, we numerically implemented the CROP and MROP models by applying each operator successively, \ie in the same fashion as illustrated in Fig.~\ref{fig:operators} and not with the batchwise summation scheme shown in Fig.~\ref{fig:acquisition}(d). 
Indeed, the dimension of our problem did not challenge the storage of all intermediate data sizes. Moreover, challenges related to the parallelism of batchwise computations are beyond the scope of the present work.

For a fair comparison in computational cost, all validations were conducted using a single Nvidia A40 GPU card with 48GB of memory.

\subsection{Performance Metrics} \label{sec:metrics}

The quality of the reconstruction by each model is evaluated through visual examination of the reconstructed and residual dirty images, along with the \emph{signal-to-noise ratio} metric, both in linear scale (SNR) and logarithmic scale (logSNR).
The computational cost will be analysed using the averaged iteration time. \\

The reconstructed images are compared to their ground truth using the SNR, in dB, defined as 
\begin{equation} \label{eq:snr}
    \text{SNR}(\wt{\bs u}, \bs u) = 20 \log_{10} \bigg( \frac{\norm{\bs u}{2}}{\norm{\bs u-\wt{\bs u}}{2}} \bigg).
\end{equation}
Given the high dynamic ranges of interest, the estimated images are shown in logarithmic scale, using the mapping
\begin{equation}
    \rlog : \bs u \mapsto \frac{\max(\bs u)}{\log_{10}(d)} \log_{10} \bigg( \frac{d }{\max(\bs u)} \bs u + 1 \bigg),
\end{equation}
where $d$ is the dynamic range of $\bs u$.
The logSNR computes the SNR of the images mapped to logarithmic scale as 
\begin{equation} \label{eq:logsnr}
    \text{logSNR}(\wt{\bs u}, \bs u) = \text{SNR}(\rlog(\wt{\bs u}), \rlog(\bs u)).
\end{equation}

For any sensing model, analogously as for the classical operator in \eqref{eq:dirty_image_im}, the dirty image is computed as
\begin{equation} \label{eq:dirty}
    \bs\sigma_\rmd := \real{\bar\imop^* \bs b}
\end{equation}
where $\bar\imop$ and $\bs b$ are defined above \eqref{eq:usara}.
The \emph{residual} dirty image is computed as 
\begin{equation} \label{eq:residual}
    \bs r := \bs\sigma_\rmd - \real{\bar\imop^* \bar\imop} \wt{\bs\sigma},
\end{equation}
where $\wt{\bs\sigma}$ is the reconstructed image. 
If the reconstruction succeeds, the residual image should only include noise back-projected by $\bar\imop^*$. 

Finally, the \emph{relative error} map $\bs\sigma_{\rm r}$ gives the relative difference between the ground truth $\bs\sigma$ and reconstructed image $\wt{\bs\sigma}$ where the pixel intensity of the ground truth $\sigma_n$ at pixel index $n$ is above the assumed noise level $\bar\tau$, and 0 elsewhere, \ie 
\begin{equation} \label{eq:rel_diff}
    (\sigma_{\rm r})_n := \left\lbrace \begin{array}{cc}
        1 - \frac{\wt{\sigma}_n}{\sigma_n}, & \text{if}~\sigma_n > \bar\tau  \\
        0, & \text{elsewhere.} 
    \end{array} 
    \right.
\end{equation}
\section{Simulation Results} \label{sec:results}

This section summarises the results of all experiments. Both quantitative results (SNR, logSNR, and timing curves), and sample image reconstructions are analysed. 

\subsection{Quantitative Results} \label{sec:fig_snr}

\begin{figure}
\centering
  \includegraphics[width=1\linewidth]{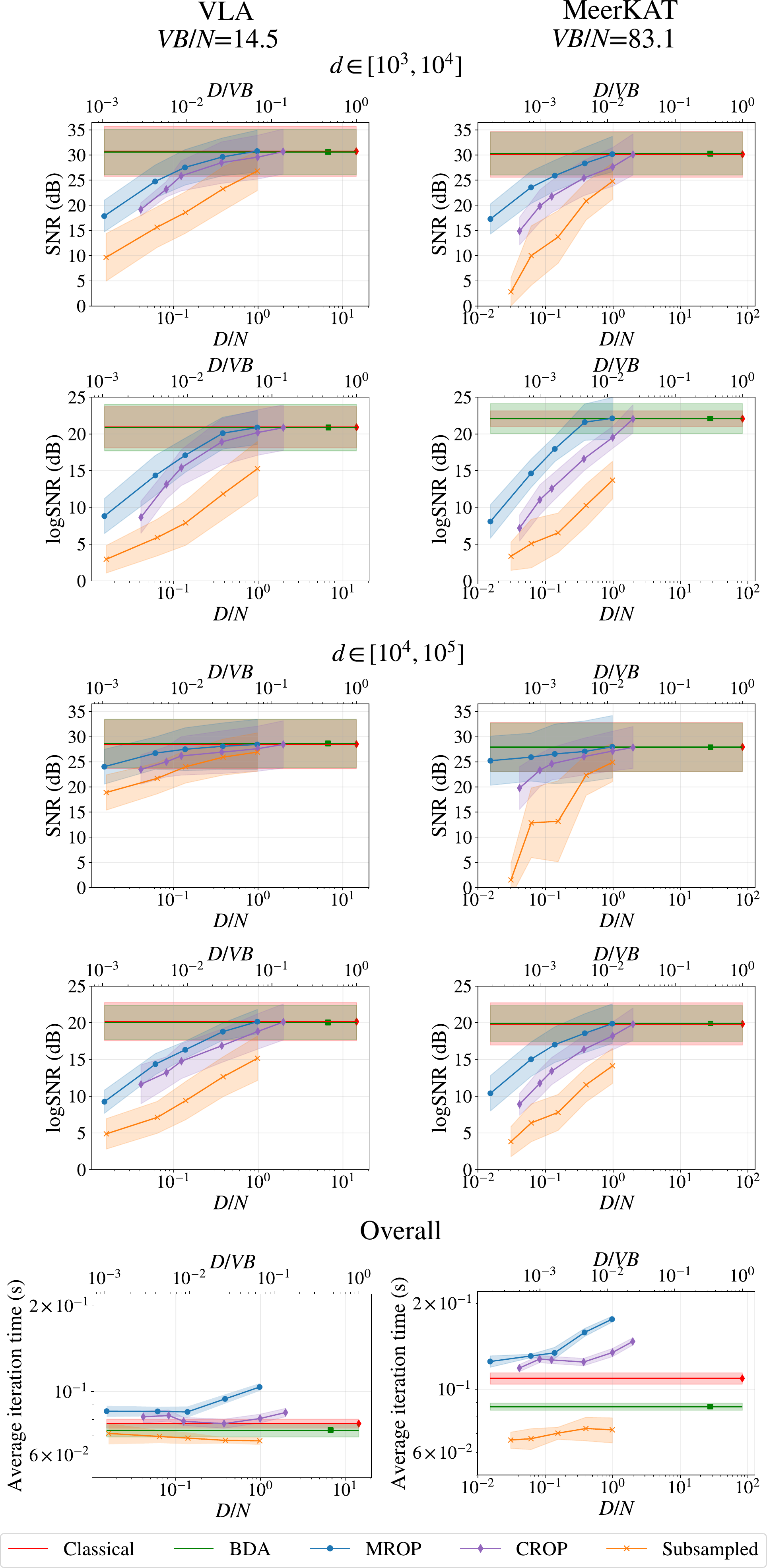}
  \caption{Reconstruction metrics: SNR and logSNR, and average iteration time (left column: VLA, right column: MeerKAT) of the classical, subsampled, BDA, CROP, and MROP models as functions of the data-to-image ratio ($D/N$, bottom x-axis) and the data-to-visibility ratio ($D/VB$, top x-axis), both in logarithmic scale, where $D=VB_{\rm sub}$ for the subsampled model, $D=VB_{\rm bda}$ for BDA, $D=P_{\rm crop}B$ for CROP, and $D=PM$ for MROP. 
  The SNR and logSNR plots are separated by the dynamic range intervals $[10^3, 10^4]$ and $[10^4, 10^5]$, while the average iteration time plots are computed with all simulations.
  \textbf{Left column: VLA.}
  The visibility-to-image ratio is $VB/N=14.5$.
  In the low (resp. high) dynamic range interval, the reconstruction qualities are (SNR, logSNR)$~=(30.73, 20.90)$ (resp. (SNR, logSNR)$~=(28.50, 20.15)$) using the reference classical model. 
  \textbf{Right column: MeerKAT.}
  The visibility-to-image ratio is $VB/N=83.1$.
  In the low (resp. high) dynamic range interval, the reconstruction qualities are (SNR, logSNR)$~=(30.13, 22.08)$ (resp. (SNR, logSNR)$~=(27.95, 19.84)$) using the reference classical model.
  }
  \label{fig:curves}
\end{figure}

\paragraph*{Overview.}

Fig.~\ref{fig:curves} provides a quantitative comparison of the image reconstruction performance in terms of SNR, logSNR, and average iteration time between the classical, subsampled, BDA, CROP, and MROP models. 
Results are reported across the range of observation settings considered, \emph{i.e.}~for VLA in a low visibility-to-image regime $VB/N=14.5$ (Fig.~\ref{fig:curves} left column) and MeerKAT in the high visibility-to-image regime $VB/N=83.1$ (Fig.~\ref{fig:curves} right column), as well as for both a low dynamic range regime $d\in[10^3, 10^4]$ (Fig.~\ref{fig:curves} rows 1 \& 2) and high dynamic range regime $d\in[10^4, 10^5]$ (Fig.~\ref{fig:curves} rows 3 \& 4). 
In all cases, model performance is reported against the data-to-image ratio $D/N$ and data-to-visibility ratio $D/VB$. 
The reconstruction metrics and times are represented as solid lines for their averaged value over the $\sim 50$ simulations in each telescope and dynamic range category and shaded region for their standard deviation. 
The classical and BDA models are represented by horizontal lines since their data sizes are constant.

\paragraph*{SNR and logSNR curves.}

Across all observation settings considered, the classical model, BDA model, CROP model with $P_{\rm crop}B\simeq 2N$, and MROP model with $PM\simeq N$ all converge to very similar SNR and logSNR metrics. 
Both the CROP and MROP metrics exhibit a progressive decrease of reconstruction quality when the number of measurements decreases from those values. 
Firstly, the alignment of BDA with the classical model is expected as averaging is by construction limited to preserve an accurate Fourier model. 
Secondly, the results confirm that compressing the visibilities into CROPs or MROPs is an efficient dimensionality reduction approach, delivering the same reconstruction quality as the classical model while effectively reducing memory requirements. 
Thirdly, these results also suggest that the image size is a reference point for the necessary number of projections. 
Fourthly, the results also point to an extra efficiency of MROP over CROP, with respective memory requirements $\cl O(N)$ and $\cl O(2N)$. 
This confirms that MROP's modulations effectively address the redundancy in the information carried by the $B$ batches of CROP projections (see Section \ref{sec:post_corr}), justifying the more involved compression setup. 
Finally, as expected, the reconstruction quality deteriorates significantly when subsampling is used for data dimensionality reduction.

\paragraph*{Timing curves.}

The average time of one uSARA iteration is reported for all models in the last row of Fig.~\ref{fig:curves}, left for VLA and right for MeerKAT.
We recall that, as mentioned in Section~\ref{sec:usara}, each uSARA iteration involves the gradient term $\real{\bar\imop^* \bar\imop \bs u - \bar\imop^* \bs b}$, and the proximal operator of the average sparsity prior. 
While the computational cost of the former is directly impacted by the sensing model, the latter is predominantly dominated by the image size. 
In this context, different sensing models and data regimes will only imprint significant variations of the average iteration times in regimes where $D>N$. 
For the classical and BDA models, the measurement operator is a NUFFT, whose cost is dominated by the computation of $\bs G$ or $\bs G_{\rm bda}$, and thus $D$, as $D>N$. 
In this context, the reported average time per iteration is larger for the classical model than for the BDA model, and larger in the high visibility-to-image regime of MeerKat than in the low visibility-to-image regime of VLA. For the subsampled model, the measurement operator is still a NUFFT, whose cost is however dominated by the FFT $\bs F$, as $D \leq N$. 
Consequently, as observed in the graphs, the average time per iteration for the subsampled model is essentially independent of $D$, and lower than for the classical and BDA models. 

For the CROP and MROP models, the measurement operator is that of the classical model, followed by the random projections, and modulations in the case of MROP. 
The graphs thus naturally reveal that the CROP and MROP iterations are slower than the classical ones. 
But overall, the CROP and MROP average iteration always remains less than twice the time of the classical model. 
In line with the complexity analysis of Section~\ref{sec:complexity}, which reveals a dependency of the computational cost in $J+2P$ (resp.~$J+2P_{\rm crop}$), the curves show a stagnation followed by  a linear increase with the number of MROPS (resp.~CROPs). 
The CROP times are also smaller than the MROP times as $P_{\rm crop}=0.6P$.

\subsection{Sample Image Reconstructions} \label{sec:im_rec}

\begin{figure*}
\centering
\includegraphics[width=0.85\linewidth]{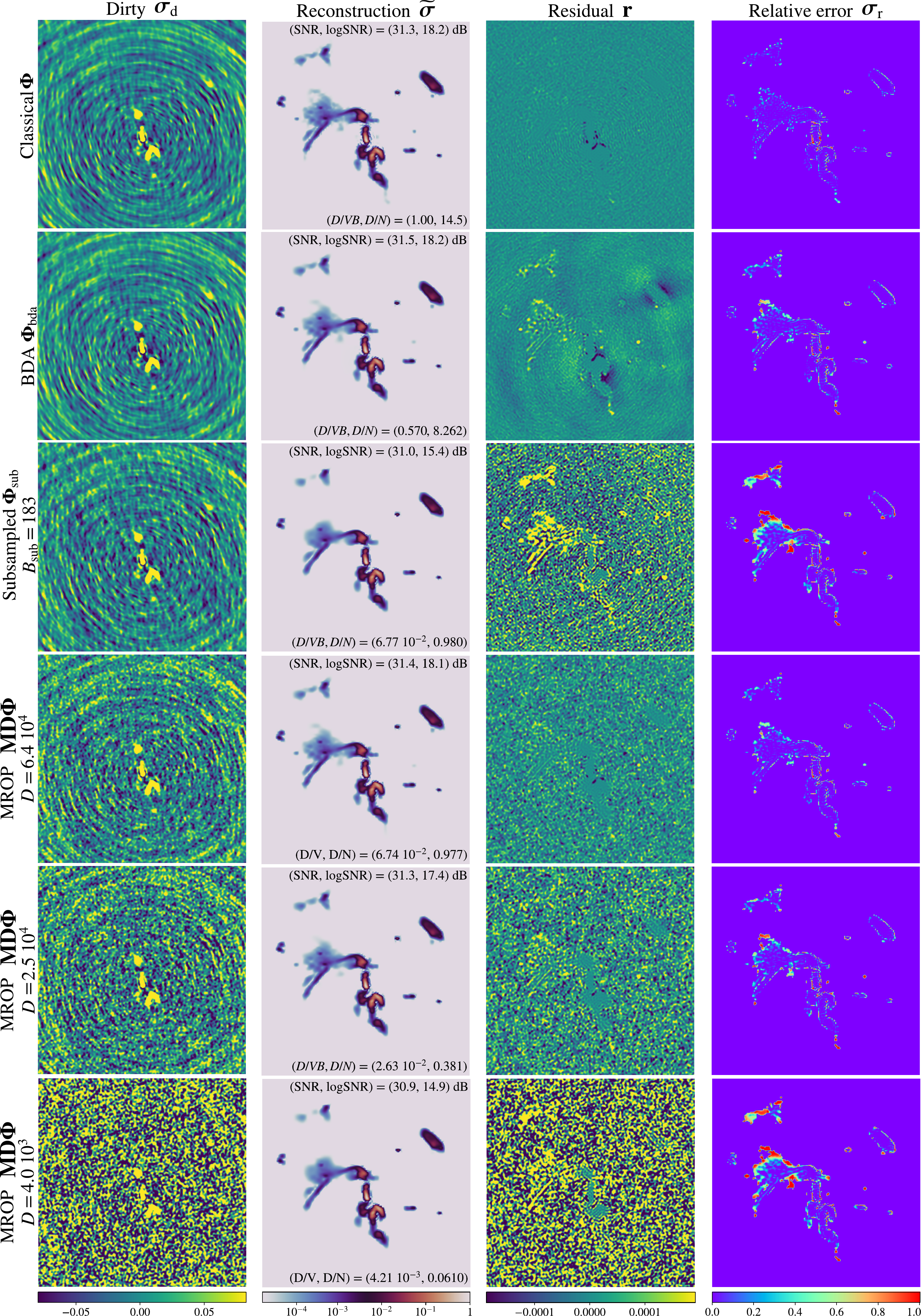}
\caption{Reconstruction results of the Abell2034 image with a dynamic range $d = 8.6~10^4$ and sampled in the regime $VB/N = 14.5$ with VLA.
Each row shows, from left to right, the dirty image $\ddirty$, image reconstruction $\wt{\bs\sigma}$ (in logarithmic scale), residual dirty image $\bs r$, and relative error map $\bs\sigma_\rmr$ for each sensing model. 
Values of the evaluation metrics (SNR, logSNR) and the compression ratios ($D/VB$, $D/N$) are reported at the top and the bottom of each reconstructed image, respectively. 
The first three rows show the results using the classical, BDA, and subsampled ($VB_{\rm sub} = 64\,233$) models, respectively.
The last three rows show the results using the MROP model with $D/N \in \{9.8~10^{-1}, 3.8~10^{-1}, 6.1~10^{-2}\}$.
}
\label{fig:abell}
\end{figure*}

\begin{figure*}
\centering
\includegraphics[width=0.85\linewidth]{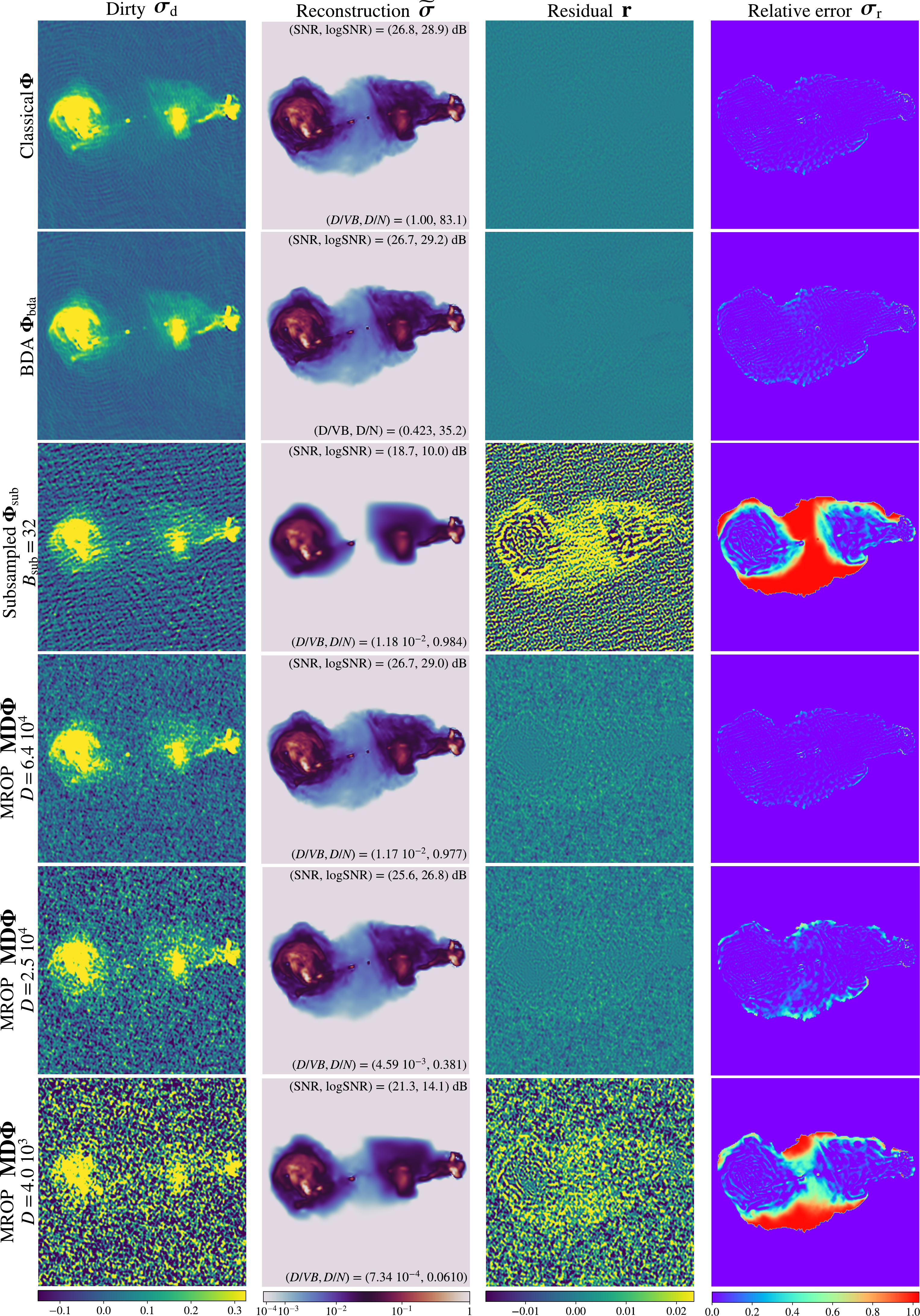}
\caption{
Reconstruction results of the 3c353 image with a dynamic range $d = 1.1~10^3$ and sampled in the regime $VB/N = 83.1$ with MeerKAT.
Each row shows, from left to right, the dirty image $\ddirty$, image reconstruction $\wt{\bs\sigma}$ (in logarithmic scale), residual dirty image $\bs r$, and relative error map $\bs\sigma_\rmr$ for each sensing model. 
Values of the evaluation metrics (SNR, logSNR) and the compression ratios ($D/VB$, $D/N$) are reported at the top and the bottom of each reconstructed image, respectively. 
The first three rows show the results using the classical, BDA, and subsampled ($VB_{\rm sub} = 64\,512$) models, respectively.
The last three rows show the results using the MROP model with $D/N \in \{9.8~10^{-1}, 3.8~10^{-1}, 6.1~10^{-2}\}$.
}
\label{fig:3c353}
\end{figure*}

\paragraph*{Overview.}

Fig.~\ref{fig:abell} displays the reconstruction results for the classical, subsampled, BDA, and MROP sensing models for a chosen inverse problem simulated by VLA using the image of the galaxy cluster Abell 2034, with dynamic range $d = 8.6~10^4$ in low visibility-to-image regime $VB/N=14.5$. The CROP reconstructions are not provided here for conciseness.
From top to bottom, the rows are arranged by the data size in a descending order, showing the results of the classical model, BDA model, subsampled model for $B_{\rm sub} = 183$, and MROP model 
with $D/N \in \{9.8~10^{-1}, 3.8~10^{-1}, 6.1~10^{-2}\}$.
From left to right, we have the dirty image $\ddirty$, reconstructed image $\wt{\bs\sigma}$ in logarithmic scale, its corresponding residual image $\bs r$, and the relative error map $\bs\sigma_{\rm r}$. 
Each reconstruction is accompanied by its (SNR, logSNR) values in dB at the top and the data-to-visibility and data-to-image compression ratios $(D/VB, D/N)$ at the bottom. 

Fig.~\ref{fig:3c353} follows exactly as Fig.~\ref{fig:abell} for a chosen inverse problem simulated by MeerKAT using the image of the radio galaxy 3c353, with dynamic range $d = 1.1~10^3$ in the $uv$ sampling regime $VB/N = 84.3$.
It differs from Fig.~\ref{fig:abell} for the subsampled model, where $B_{\rm sub}=32$.

\paragraph*{Analysis of Figs.~\ref{fig:abell} and \ref{fig:3c353}.}

Overall, three out of the six results shown in Fig.~\ref{fig:abell}, namely those by the classical and BDA models and the MROP model with $D/N \simeq 1$, perform similarly to one another.
The reconstruction by the classical model can be deemed the most faithful to the ground-truth image in Fig.~\ref{fig:groundtruth}b, since its residual dirty image is the closest to be noise-like and its relative error map is the faintest among all models. 
The reconstructed images and relative error maps of the BDA model and the MROP model with $D/N \simeq 1$ are almost visually indistinguishable, notwithstanding the data size of the latter being one order of magnitude smaller than the former. 
It should be noted that the MROP model with $D/N = 3.8~10^{-1}$ performs marginally worse than the MROP model with $D/N \simeq 1$, as evidenced by a drop of $0.1$ dB and $0.7$ dB in the SNR and logSNR values respectively, and loss of some faint emission in the reconstructed image by the former.
The reconstructed images of both the subsampled model and the MROP model with $D/N = 6.1~10^{-2}$ struggle to recover the faint emission, which is evidenced by much higher residuals and relative errors and much lower logSNR values.

The same conclusions hold for Fig.~\ref{fig:3c353}, despite very different $uv$-coverage and dynamic range regimes. 
In this case, a more significant difference is observed in the reconstruction quality of the MROP model with $D/N \simeq 1$ and $D/N = 3.8~10^{-1}$, where the latter exhibits loss of details in faint emission and a decrease of $1.1$ dB and $2.2$ dB in SNR and logSNR, respectively. Furthermore, while both metrics are significantly inferior to the reconstructions, the MROP model with $D/N = 6.1~10^{-2}$ outperforms the subsampled model in this example both visually and quantitatively.

These results corroborate the quantitative results of Section~\ref{sec:fig_snr} with MROP enabling a significant reduction of data dimensionality to around the image size while preserving reconstruction quality.
\section{Real Data Proof Of Concept} \label{sec:real_data}

This section studies the performance of the MROP model on real data obtained from VLA observations in comparison with the classical model, using the image reconstruction algorithm uSARA.

\subsection{Observation and Imaging Settings}
The acquired data correspond to observations of the quasar 3C 273 at frequency 8.872 GHz in the X band using configuration A\footnote{\url{https://public.nrao.edu/telescopes/vla/}.} of the VLA, with a single pointing centred at the inner core of 3C 273, given by the coordinates right ascension $12\textrm{h}29\textrm{m}6.700\textrm{s}$~(J2000) and declination $+2^{\textrm{o}} 3{\arcmin} 8.60{\arcsec}$. The accumulated observation duration is approximately 10.44 hr, conducted over 6 days and acquired with an integration time step of $IT=1$s. The data have already been carefully calibrated for direction-independent effects (DIEs).

We consider the formation of an image of size $N = 320 \times 320$, with a pixel size of about $0.0634$ arcsec, corresponding to a super-resolution factor of 1.5, from uniformly-weighted data. The data size of the classical model is $VB = 1.13~10^6$, corresponding to a data-to-image ratio of $D/N = 11.1$, similar to the VLA simulations considered in Sections~\ref{sec:setting}-\ref{sec:results}. For the MROP model, we consider a set of $PM$ values, including $PM = 10^5$, which corresponds to a data-to-image ratio of $D/N=9.8~10^{-1}\simeq N$, as well as lower values.

\subsection{MROP imaging in the presence of flagged visibilities}
In a real observation setup, part of the visibilities are commonly flagged due to radio frequency interference (RFI) or other instrumental effects. In the classical imaging model, these are simply discarded from the Fourier measurement operator model (see Fig.~\ref{fig:acquisition}(a)). In the MROP imaging model, as full covariance matrices must be formed to support random projection (see Fig.~\ref{fig:acquisition}(d)), flagged data cannot just be discarded. However, they can simply be zeroed in the covariance matrix $\bsmat C_b$ before random-one projection in \eqref{eq:31_sample_rop}, so that their contribution cancels out in the multiplications with the $\bs\alpha$ and $\bs\beta$ vectors. The MROP imaging model thus trivially generalises to address imaging in the presence of flagged visibilities. 
This strategy was used to account for flagged visibilities in the real data at hand.

We note however that further considerations are warranted to allow for the flagging process during MROP acquisition, as visibilities are not explicitly accessible after acquisition. While such considerations lie beyond the scope of the current work, we highlight that this would typically require real-time flagging and zeroing, either at the level of antenna voltages in the context of antenna-based MROP acquisition (see Fig.~\ref{fig:acquisition}(b)), or at the level of the visibilities in the context of visibility-based MROP acquisition (see Fig.~\ref{fig:acquisition}(c)).

\subsection{Results}
\begin{figure*}
\centering
\includegraphics[width=0.95\linewidth]{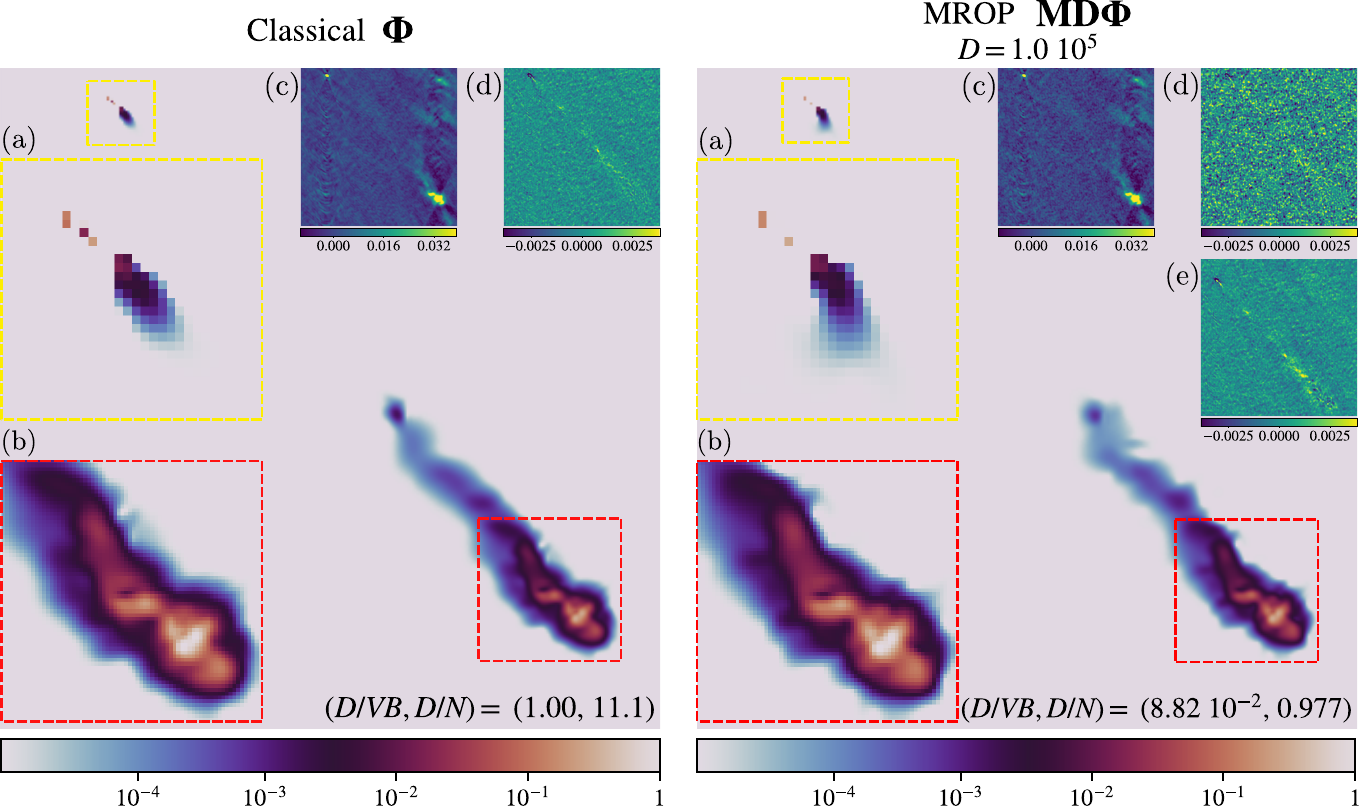}
\caption{Reconstructions of VLA real the quasar 3C 273. The sampling regime is such that $VB/N = 11.1$. From left to right are the respective reconstructions of the classical and MROP ($D/N=9.8~10^{-1}$) models, displayed in logarithmic scale. Each reconstructed image is overlaid by (a) a zoom on the centre of the inner core, (b) a zoom on the end of the jet, (c) the corresponding dirty image and (d) the corresponding residual dirty image. The MROP reconstructed image is additionally overlaid by (e) the residual dirty image computed using the classical measurement operator.
Values of the compression ratios ($D/VB$, $D/N$) are reported at the top of each reconstructed image.
}
\label{fig:3c273}
\end{figure*}

Fig.~\ref{fig:3c273} displays the uSARA reconstruction results from the 3C 273 data for both the classical model (left) and MROP model (right) at data-to-image ratio $D/N = 9.8~10^{-1}$. 
Visual inspection of 3C 273 reconstructions reveals that the imaging quality achieved with the MROP model is consistent with that of the classical model. 
Both reconstructions exhibit nearly identical peak intensity values of 0.07 and recover features spanning 4 orders of magnitude. 
The reconstructions from both models achieve high data fidelity, as evidenced by their noise-like residual dirty images (Fig.~\ref{fig:3c273}, left and right panels (d)).
Fig.~\ref{fig:3c273}, right panel (e), shows the residual dirty image associated with the MROP reconstruction, computed using \eqref{eq:residual} with the dirty image and measurement operator of the classical model.
This serves as a direct comparison to the residual dirty image of the classical model reconstruction (Fig.~\ref{fig:3c273}, left panel (d)), with both images appearing highly comparable and only subtle differences observed at the end of the jet. 
Moreover, the highly comparable standard deviations of these residual maps ($6.49~10^{-4}$ for the classical model and $6.87~10^{-4}$ for the MROP model) further confirm the agreement between the reconstructions.

We finally note that the additional MROP reconstructions at $D < N$ (not shown here) indicate a degradation in reconstruction quality,  consistent with the simulation results in Section~\ref{sec:results}.

\section{Conclusion and Perspectives} \label{sec:conclusion}

This paper introduced the MROP model as an efficient data compression framework for RI, detailing its computational cost and memory requirements for both the acquisition and imaging stages. 
The study explored antenna- and visibility-based approaches for acquiring MROPs. Following from a detailed theoretical complexity analysis, numerical experiments demonstrated the effectiveness of the MROP model by reconstructing realistic high dynamic range images with faint and diffuse emission across the field view in both VLA and MeerKAT observation settings, confirming its advantages over the classical, subsampled, and BDA models, and the simpler CROP model.
Moreover, experimental results on real data from VLA observations demonstrated that the MROP model achieves image reconstructions comparable to those of the classical model, even at high compression ratios.
\\

Our contributions open several perspectives for future research. 
Firstly, future research should study the combination of MROP with weighting schemes beyond uniform weighting, from natural or Briggs \citep{boone13} weightings, to \emph{adaptive} weighting \citep{yatawatta14}, to alternative preconditioning methods \citep{onose17}. 
Secondly, the theoretical recovery guarantees in \citet{leblanc2024b} only holds for the IROP model and for a simplistic sparse image model, calling for further theoretical developments to support the MROP approach and for more complex RI image models. 
This is key to providing a capability to determine \emph{before observation} the number of MROPs necessary to an optimal image reconstruction quality. 
Our simulation results suggest that the image size is a reference point for the necessary number of MROPs, which represents a driver to derive theoretical recovery guarantees. 
Thirdly, the option of precomputing the forward operator to reduce its computational cost at the price of increased memory storage opens many perspectives on the complexities discussed in Section~\ref{sec:complexity}. 
Fourthly, while this paper focused on single-frequency monochromatic intensity imaging, MROP should be generalised to wideband polarisation imaging, with a significant compression potential expected along the spectral dimension.
Fifthly, addressing the problem of joint self-calibration and imaging in the context of MROP is paramount to its potential adoption. 
In this regard, we emphasised that MROP is self-calibration friendly, with DIEs/DDEs to be introduced as additional variables before the random projections in the measurement model, thus opening the door to leveraging the ``joint self-calibration and imaging'' extension of uSARA  \citep{repetti2017non,dabbech2021cygnus}.
Sixthly, incorporating visibility flagging into MROP acquisition is no less important, for which real-time flagging approaches are to be developed.
Seventhly, the MROP approach should be validated for a variety of state-of-the-art image reconstruction algorithms beyond uSARA, from CLEAN to its most recent learned variant R2D2 
\citep{dabbech2024cleaning,aghabiglou24}. 
Finally, efficient batchwise implementations of the MROP measurement operator are warranted to enable image formation from MROP data at the large scales of interest for modern radio arrays.

\section*{Acknowledgment}

The authors warmly thank R. A. Perley (NRAO, USA) for providing the VLA observations of 3C 273, and A. Dabbech (HWU, UK) for her help in 3C 273 reconstruction result analysis. The authors acknowledge the use of the HWU high-performance computing (HPC) facility (DMOG) and associated support services, where all experiments were conducted on its HPC cluster, utilising all 32 cores of an AMD EPYC 7513 processor.
This project has received funding from FRS-FNRS (QuadSense, T.0160.24). 
The research of YW was supported by the UK Research and Innovation under the EPSRC grant EP/T028270/1 and the STFC grant ST/W000970/1. The National Radio Astronomy Observatory is a facility of the National Science Foundation operated under cooperative agreement by Associated Universities, Inc.

\section*{Data and code availability}

Python codes supporting the MROP measurement operator and its integration into uSARA will be made available as part of a future release of the \href{https://basp-group.github.io/BASPLib/}{BASPLib} code library on GitHub. BASPLib is developed and maintained by the Biomedical and Astronomical Signal Processing Laboratory (\href{https://basp.site.hw.ac.uk/}{BASP}).

\bibliographystyle{mnras}
\bibliography{biblio} 

\bsp	
\label{lastpage}
\end{document}